\documentclass[12pt]{article}
\usepackage[sort&compress,square,comma,numbers]{natbib}
\usepackage{amsmath,amssymb,graphicx,slashed,mathtools,bm}
\usepackage{feynmp}
\usepackage{subfigure}
\usepackage{multirow}
\usepackage{hhline}
\usepackage{makecell}

\usepackage[
	colorlinks=true,
	citecolor=black,
	linkcolor=black,
	urlcolor=blue,
	hypertexnames=false]{hyperref}

\makeatletter
\def\endfmffile{%
	\fmfcmd{\p@rcent\space the end.^^J%
		end.^^J%
		endinput;}%
	\if@fmfio
	\immediate\closeout\@outfmf
	\fi
	\ifnum\pdfshellescape>\z@
	\immediate\write18{mpost \thefmffile}%
	\fi}
\makeatother

\newcommand{\arXiv}[2]{\href{http://arxiv.org/pdf/#1}{{\tt #2/#1}}}
\newcommand{\arXivold}[1]{\href{http://arxiv.org/pdf/#1}{{\tt #1}}}

\newcommand{\beq}{\begin{eqnarray}}
\newcommand{\eeq}{\end{eqnarray}}

\def\tilde#1{\widetilde{#1}}

\begin{document}
\vspace*{-4cm}
\begin{flushright}
UCI-TR-2020-12
\end{flushright}
\vspace{1.5cm}
\begin{center} 
{\huge \bf Spurious Poles in the Scattering of}
\\ 
\vskip 8pt
{\huge \bf Electric and Magnetic Charges} 
\end{center}

\begin{center} 

{\bf John Terning}$^1$ and {\bf Christopher B. Verhaaren}$^2$ \\
\end{center}
\vskip 8pt
\begin{center} 
$^1${\it Center for Quantum Mathematics and Physics (QMAP)\\Department of Physics, University of California, Davis, CA 95616}\\
$^2${\it Department of Physics and Astronomy, University of California, Irvine, USA}
\end{center}

\vspace*{0.1cm}
\begin{center} 
{\tt 
 \href{mailto:jterning@gmail.com}{jterning@gmail.com}\,
 \href{mailto:cbverhaaren@ucdavis.edu}{cverhaar@uci.edu}}

\end{center}

\centerline{\large\bf Abstract}
\begin{quote}
Theories with both electric and magnetic charges (``mutually non-local" theories) have several major obstacles to calculating 
scattering amplitudes. Even when the interaction arises through the kinetic mixing of two, otherwise independent, $U(1)$'s, so that
all low-energy interactions are perturbative, difficulties remain: using a self-dual, local formalism leads to spurious poles at any finite order in perturbation theory. Correct calculations must show how the spurious poles cancel in observable scattering amplitudes. Consistency requires that one type of charge is confined as a result of one of the $U(1)$'s being broken. Here we show how the constraints of confinement and parity conservation on observable processes manages to cancel the spurious poles in scattering and pair production amplitudes, paving the way for systematic studies of the experimental signatures of ``dark" electric-magnetic processes. Along the way we demonstrate some novel effects in electric-magnetic interactions, including that the amplitude for single photon production of magnetic particles by electric particles vanishes.
\end{quote}


\section{Introduction\label{s.Intro}}
Seventeen years after Dirac discovered charge quantization \cite{Dirac} he returned to the theory of electric and magnetic charges, showing that a Lorentz 
invariant Lagrangian description is necessarily non-local \cite{Dirac2}. Later Zwanziger \cite{Zwanziger:1971} discovered how to write a local Lagrangian using a second, magnetic photon field, but the constraint that reduces the degrees of freedom to that of the on-shell photon requires a spacelike Lorentz violating four-vector in the Lagrangian.  This four-vector is associated with the direction of the Dirac string, which is itself merely an unphysical gauge artifact. After the discovery of $SL(2,\mathbb{Z})$ duality \cite{Cardy1,Vafa,Witten2,Lozano} it was shown that Zwanziger's Lagrangian is actually self-dual \cite{Csaki:2010rv}. (An alternative self-dual formulation was uncovered much later, independently, by Schwarz and Sen \cite{Schwarz:1993vs}, extending work by Henneaux and Teitelboim \cite{Henneaux:1988gg}, which requires a timelike Lorentz violating four-vector~\cite{Maznytsia:1998xw}.\footnote{We show in Appendix~\ref{a.twoPot} that the Schwarz-Sen Lagrangian leads to a two-point function connecting electric and magnetic charges which exactly agrees with the results obtained by Weinberg (who didn't rely on a Lagrangian). Weinberg's gauge choice is related to a timelike four-vector which is  associated with Schwarz-Sen vector.}) These results imply that we cannot have a Lagrangian that has both manifest Lorentz invariance and manifest locality, we must choose one or the other. But this is simply a problem with Lagrangians, we could still hope that amplitudes are both Lorentz invariant and local.
This hope seemed to be dashed by the Weinberg paradox: Weinberg \cite{Weinberg:1965rz} showed that the one-photon exchange amplitude between an electric and a magnetic charge is not Lorentz invariant! 

Recently it was observed that Weinberg's amplitude is also not gauge invariant \cite{Terning:2018udc}, since electric charges couple with strength $e$ and magnetic charges couple with strength $4\pi/e$; an all orders resummation is required for a gauge
invariant result. In the soft-photon limit such a resummation can be performed, and it was found \cite{Terning:2018udc} that when using the local, self-dual formalism the Lorentz violating vector only contributes to a phase, so squares of scattering amplitudes are both Lorentz invariant and local.  Furthermore, the Lorentz violating vector only appears inside a topological number \cite{Terning:2018udc}, and when Dirac-Schwinger-Zwanziger charge quantization \cite{Dirac,Schwinger,Zwanziger:1968rs}  is imposed the phase is always a multiple of $2 \pi$, so the phase cannot be detected even in interference experiments. There has also been further work using on-shell amplitudes \cite{Caron-Huot:2018ape,Moynihan:2020gxj,Csaki:2020inw}.

A parallel set of developments \cite{Holdom:1985ag,Brummer:2009cs,Bruemmer:2009ky,Sanchez:2011mf,Hook:2017vyc,Terning:2018lsv} occurred in the study of $U(1)$ theories with kinetic mixing. If a photon coupled to electric charges is kinetically mixed to a massive ``dark" photon coupled to ``dark"  magnetic charges, then the low-energy theory (below the ``dark" photon mass\footnote{In what follows we continue the common practice of referring to quantities related to the new $U(1)$ as dark, without scare quotes.}) has both electric and magnetic charges, but charge quantization does not apply, since the magnetic charges are suppressed by the size of the kinetic mixing. This leads to a new form of Weinberg paradox. In this case the magnetic charge can be parametrically small and Weinberg's calculation should be a very good approximation to the actual amplitude. The confusion is further compounded by the fact that photon propagators that connect electric and magnetic charges contain unphysical poles involving the Lorentz violating four-vector. The proper treatment of these poles is a necessary first step toward the study of these perturbative electric-magnetic interactions.

The origin and form of the spurious pole can be easily seen by relating the field strength $F^{\mu\nu}$ and its Hodge dual ${}^\ast \!F^{\mu\nu}=\frac12\varepsilon^{\mu\nu\alpha\beta}F_{\alpha\beta}$. Suppose we write 
\beq
F_{\mu\nu}=\partial_\mu A_\nu-\partial_\nu A_\mu, \ \ \ \ {}^\ast F_{\mu\nu}=\partial_\mu \widetilde{A}_\nu-\partial_\nu \widetilde{A}_\mu,
\eeq
where $A_\mu$ is the usual gauge potential representing the photon field, which has a local coupling to electric currents and $\widetilde{A}_\mu$ is the dual photon field, which has a local coupling to magnetic currents. Then, in momentum space, the relation between the photon and the dual photon can be written as
\beq
p_\mu A_\nu-p_\nu A_\mu=\varepsilon_{\mu\nu\alpha\beta}p^\alpha \widetilde{A}^\beta.
\eeq
If we contract a vector $n^\mu$ into both sides we can then ``solve" for $A_\nu$ as
\beq
A_\nu-p_\mu \frac{n\cdot A}{n\cdot p}=\frac{\varepsilon_{\mu\nu\alpha\beta}}{ n\cdot p}n^\mu p^\alpha \widetilde{A}^\beta~,\label{e.mutnonloc}
\eeq
where we recognize that the second term on the left-hand side has the form of a gauge transformation $A_\mu\to A'_\mu=A_\mu-\partial_\mu f$. In particular, it vanishes when we pick the gauge in which $n\cdot A=0$. This is the essence of mutual non-locality. When both electric and magnetic charges are present neither the $A_\mu$ nor $\widetilde{A}_\mu$ form of the photon field can be used without introducing these nonlocal $1/(n\cdot p)$ factors in some interactions. In this paper we show how a restriction on the set of physical amplitudes ensures a cancellation of these unphysical poles. The key observation is that the construction of a low-energy theory without charge quantization relies on a dark photon mass, which implies that magnetic charge is confined \cite{Artru:1976qw} by Nielsen-Olesen flux tubes \cite{Nielsen:1973cs}. In this case $n^\mu$ is associated with a physical string of dark magnetic flux. This makes it reasonable for physical quantities to depend on $n^\mu$, but not that the orientation of the string can produce poles in scattering processes.
We find that the restriction to asymptotic states with confined  magnetic charges is sufficient to cancel the spurious poles.

In the following section we review Zwanziger's self-dual formalism, explaining how physical, perturbative magnetic couplings arise through the kinetic mixing of two $U(1)$ gauge theories. We then clarify 
the action of discrete symmetries in theories with magnetic charges. In section~\ref{s.scat} we investigate $t$-channel scattering of electric and magnetic particles. We find it is essential to include the effects of the electric particle scattering off both constituents of the magnetic bound state. When this is done all spurious poles cancel when $n^\mu$ is associated with the relative position of the two magnetic particles. We also find that the magnetic charge radius, as seen by electric probes, is zero. Section \ref{s.Prod} explores $s$-channel production of magnetic pairs by the annihilation of electric particles. We find that in amplitudes for producing bound states that match the $J^{PC}$ of a single photon the  spurious pole cancels. Furthermore we prove that the  amplitude for electric particle annihilation through a single photon to pair produce magnetic monopoles vanishes. We then show that the non-vanishing photon fusion production amplitude is free of spurious poles, if $n^\mu$ is again associated with the relative position of the monopoles, as expected for the direction of a physical string. After presenting  our conclusions 
we provide more details about the propagators for the gauge fields in two-potential Lagrangians and the $SL(2,\mathbb{Z})$ duality structure of Zwanziger's formalism for both spin zero and spin half matter in Appendices~\ref{a.twoPot} and~\ref{a.ScalDual}. The formalism for relativistic spin projection matrices is reviewed in Appendix~\ref{a.RelProj}.

\section{Self-Dual Lagrangian with Perturbative Charges\label{s.ZLag}}
This section briefly reviews Zwanziger's self-dual formalism \cite{Zwanziger:1971}, focusing on the case of perturbative electric and magnetic charges. For simplicity we only consider $CP$ invariant theories. Further demonstrations of the self-dual nature of the Lagrangian for fermionic and scalar matter are given in Appendix~\ref{a.ScalDual}. The simplest self-dual Lagrangian is given by
\begin{align}
\mathcal{L}_Z=&-\frac{n^\alpha n^\mu}{8\pi n^2}g^{\beta\nu}\frac{4\pi}{e^2}\left(F^A_{\alpha\beta}F^A_{\mu\nu}+F^B_{\alpha\beta}F^B_{\mu\nu} \right)+\frac{n^\alpha n_\mu}{16\pi n^2}\varepsilon^{\mu\nu\gamma\delta} \frac{4\pi}{e^2}\left( F^B_{\alpha\nu}F^A_{\gamma\delta}-F^A_{\alpha\nu}F^B_{\gamma\delta}\right)\nonumber\\
&-A_\mu J^\mu-\frac{4\pi}{e^2}B_\mu K^\mu~,\label{e.zLag}
\end{align}
where $g_{\alpha\beta}=\text{diag}(1,-1,-1,-1)$ is the Minkowski metric and we have used the notation
\beq
F^X_{\mu\nu}=\partial_\mu X_\nu-\partial_\nu X_\mu~.
\eeq
The gauge potentials $A_\mu$ and $B_\mu$ have local couplings to the conserved electric and magnetic currents,  $J^\mu$ and $K^\mu$ respectively, which are fermion bilinears. Charged scalars lead to the same types of results, as shown in Appendix~\ref{a.ScalDual}, even though they cannot be written in this simple form. The spacelike vector $n^\mu$ plays the role of the direction of the Dirac string \cite{Terning:2018lsv}. 

This self-dual Lagrangian leads to photon propagators that describe interactions between two electric charges (details of the derivation are given in Appendix~\ref{a.twoPot} )
\beq
\Delta^{\mu\nu}_{AA}=-\frac{i}{k^2}\left(g^{\mu\nu}-\frac{k^\mu n^\nu+k^\nu n^\mu}{n\cdot k} \right),
\eeq
where the $n^\mu$ dependance does not contribute to physical amplitudes due to the Ward-Takahashi  identity for current conservation and we have dropped a gauge fixing dependent term that similarly vanishes in amplitudes. The form of the propagator between two magnetic charges, $\Delta^{\mu\nu}_{BB}$,  is identical. However, the mixed charge propagator manifests the mutual ``non-locality" of electric and magnetic interactions. In agreement with Weinberg's result~\cite{Weinberg:1965rz}, and the expectations from Eq.~\eqref{e.mutnonloc}, this propagator has the form\footnote{This agrees with Weinberg's result, but is not quite equal to it. The difference is due to Weinberg's $n^\mu$ being timelike, while Zwanziger's is spacelike. Appendix~\ref{a.twoPot} shows that the Schwarz-Sen Lagrangian, a self-dual two potential Lagrangian with timelike $n^\mu$, reproduces Weinberg's result exactly.}
\beq
\Delta^{\mu\nu}_{AB}=\frac{i}{k^2}\frac{\varepsilon^{\mu\nu\alpha\beta}n_\alpha k_\beta}{n\cdot k}~.\label{e.mixedProp}
\eeq
It is this spurious pole in $n\cdot k$ which must be cancelled in all physical amplitudes.

Zwanziger's self-dual Lagrangian was constructed precisely so that the Euler-Lagrange equations lead to the usual Maxwell equations 
\beq
\partial_\nu F^{\mu\nu}=e^2J^\mu, \ \ \ \ \partial_\nu {}^\ast\! F^{\mu\nu}=4\pi K^\mu~,\label{e.MaxEq}
\eeq
with
\begin{align}
F_{\mu\nu}=&\frac{n^\alpha}{n^2}\left(n_\mu F^A_{\alpha\nu}-n_\nu F^A_{\alpha\mu}-\varepsilon_{\mu\nu\alpha\beta}n_\gamma F^{B\gamma\beta} \right),\label{e.F}\\
{}^\ast\! F_{\mu\nu}=&\frac{n^\alpha}{n^2}\left(n_\mu F^B_{\alpha\nu}-n_\nu F^B_{\alpha\mu}+\varepsilon_{\mu\nu\alpha\beta}n_\gamma F^{A\gamma\beta} \right).\label{e.Fstar}
\end{align}
Notice that these tensors satisfy the definition of  the Hodge dual: ${}^\ast\! F_{\mu\nu}=\frac12\varepsilon_{\mu\nu\alpha\beta}F^{\mu\nu}$.

Next, we include a separate dark sector (with fields labeled by a subscript $D$) with a small 
kinetic mixing \cite{Holdom:1985ag,Terning:2018lsv} between the visible and dark $U(1)$ field strengths:
\beq
\frac{\varepsilon}{2}  \,F_{\alpha\beta}F^{\alpha\beta}_D=\varepsilon \frac{n^\alpha n^\mu}{n^2}g^{\beta\nu}\left(F^A_{D\alpha\beta}F^A_{\mu\nu}-F^B_{D\alpha\beta}F^B_{\mu\nu} \right)~.
\label{mixing}
\eeq
If the dark photon also has a mass, through the usual Higgs mechanism, then there is a unique basis that diagonalizes the interactions.
The dark photon mass can arise from either a dark electric or dark magnetic condensate \cite{Terning:2018lsv}. For simplicity we only consider the case of a dark electric condensate with mass term
\beq
{\mathcal L}_{{\rm mass}}=\frac{m_{D}^2}{2}A_{D\mu} A_D^\mu~.\label{e.Emass}
\eeq
We also assume there are no light fields with dark electric charges or visible magnetic charges. In this case the currents that couple to the diagonalized gauge potentials, denoted with a bar, are given by
\begin{align}
\overline{J}_\mu=&~J_\mu, &\overline{J}_{D\mu}=&~\varepsilon \frac{e}{e_D} J_\mu,\nonumber\\
\overline{K}_\mu=& -\varepsilon \frac{e}{e_D}  K_{D\mu}, &\overline{K}_{D\mu}=&~K_{D\mu}.\label{e.DiagCurrents}
\end{align}
For small $\varepsilon$, $e$, and $e_D$ we can have a weakly coupled theory with both electric and magnetic charges.
Below the mass of the dark photon we just have a perturbative $U(1)$ theory with both electric and magnetic charges.
Since this theory violates Dirac-Schwinger-Zwanziger \cite{Dirac,Schwinger,Zwanziger:1968rs} charge quantization  it is clear that Dirac's demonstration of the non-observability of the 
string no longer holds. In fact the dark photon mass we have assumed implies that dark magnetic charges are confined \cite{Artru:1976qw}, as originally argued by `t Hooft and Mandelstam \cite{'tHooftMandelstam}, and the unobservable Dirac string
is replaced by a Nielsen-Olesen flux tube \cite{Nielsen:1973cs}. While this gives a physical meaning to $n^\mu$ in the mixed charge propagator in Eq.~\eqref{e.mixedProp}, the spurious $n\cdot k$ pole is still problematic. As we show in the following sections, when the confined nature of the magnetic charges is taken into account the $n\cdot k$ poles cancel in physical amplitudes. 

While the form of the couplings in Eq.~\eqref{e.DiagCurrents} is required by the duality structure of $U(1)$ gauge theories \cite{Terning:2018lsv}, not all electric interaction intuition carries over to magnetic interactions. For instance, consider decoupling the dark photon by taking its mass to be parametrically large and then integrating the dark photon out of the theory. In this case it is clear that the interactions between the visible matter and the dark photon vanish, simply because the dark photon is absent. However, in this limit we seem to retain a coupling between the visible photon and the dark monopoles, in apparent contradiction to the decoupling theorem \cite{Appelquist:1974tg}. To better understand this we can examine the interaction between dark and visible particles in the original basis, with the kinetic mixing taken into account as a perturbative interaction between the two photons. 

To do this we must employ the propagators for massive gauge bosons~\cite{Balachandran:1974nw} arising from the electric mass in Eq.~\eqref{e.Emass}. The derivation of the required propagators:
\begin{align}
\Delta_{A_DA_D}^{\mu\nu}=&\frac{-i\, g^{\mu\nu}}{k^2-m_D^2}~,\\
\Delta_{B_DB_D}^{\mu\nu}=&\frac{-i}{k^2-m_D^2}\left[g^{\mu\nu}-\frac{m_D^2}{\left(n\cdot k \right)^2}\left(n^2g^{\mu\nu}-n^\mu n^\nu \right)\right],\\
\Delta_{A_DB_D}^{\mu\nu}=&\frac{i\,\varepsilon^{\mu\nu\alpha\beta}n_\alpha k_\beta}{\left(k^2-m_D^2\right)\left(n\cdot k \right)}~,
\end{align}
is given in Appendix~\ref{a.twoPot} and we have dropped $k^{\mu,\nu}$ terms that vanish when dotted into conserved currents. The double $n\cdot k$ pole in the $BB$ propagator appears concerning, but as seen below it is intimately connected to the single pole in the $AB$ propagator. In fact this extra pole is related to the linear term in the confining potential between the monopoles~\cite{Gubarev:1998ss,Chernodub:1999tv,Zakharov:1999gn}. 

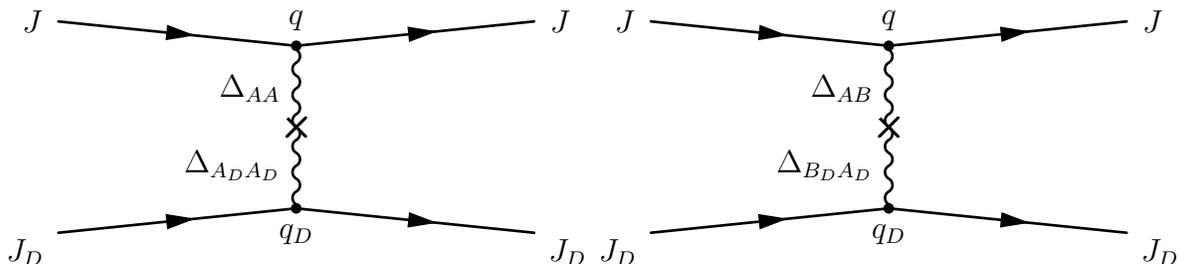
\begin{figure}[ht]
\centering
\begin{fmffile}{JJmix1}
\begin{fmfgraph*}(180,80)
\fmfpen{1.0}
\fmfstraight
\fmfleft{i1,p1,i2}\fmfright{o1,p2,o2}
\fmfv{l= $J_D$}{i1}\fmfv{l=$J_D$}{o1}
\fmfv{l= $J$,l.a=180}{i2}\fmfv{l=$J$,l.a=0}{o2}
\fmf{fermion,tension=1}{i1,v1,o1}
\fmf{fermion,tension=1}{i2,v2,o2}
\fmf{boson, tension=0.6,label=$\Delta_{A_DA_D}$,label.side=left}{v1,v}
\fmf{boson, tension=0.6,label=$\Delta_{AA}$,label.side=left}{v,v2}
\fmfv{decor.shape=circle,decor.filled=full,decor.size=1.5thick,l=$q_D$,l.a=-90}{v1}
\fmfv{decor.shape=circle,decor.filled=full,decor.size=1.5thick,l=$q$,l.a=90}{v2}
\fmfv{decor.shape=cross,decor.size=10,l.a=-25}{v}
\end{fmfgraph*}
\end{fmffile}\hspace{1cm}
\begin{fmffile}{JJmix2}
\begin{fmfgraph*}(180,80)
\fmfpen{1.0}
\fmfstraight
\fmfleft{i1,p1,i2}\fmfright{o1,p2,o2}
\fmfv{l= $J_D$}{i1}\fmfv{l=$J_D$}{o1}
\fmfv{l= $J$,l.a=180}{i2}\fmfv{l=$J$,l.a=0}{o2}
\fmf{fermion,tension=1}{i1,v1,o1}
\fmf{fermion,tension=1}{i2,v2,o2}
\fmf{boson, tension=0.6,label=$\Delta_{B_DA_D}$,label.side=left}{v1,v}
\fmf{boson, tension=0.6,label=$\Delta_{AB}$,label.side=left}{v,v2}
\fmfv{decor.shape=circle,decor.filled=full,decor.size=1.5thick,l=$q_D$,l.a=-90}{v1}
\fmfv{decor.shape=circle,decor.filled=full,decor.size=1.5thick,l=$q$,l.a=90}{v2}
\fmfv{decor.shape=cross,decor.size=10,l.a=-25}{v}
\end{fmfgraph*}
\end{fmffile}
\caption{\label{f.JJmix} The two diagrams describing the interaction between visible and dark electric currents. Scattering proceeds through $A$-$A_D$ mixing (left) and $B$-$B_D$ mixing (right).}
\end{figure}

We begin with the more familiar interaction between a visible electric charge and a dark electric charge. If we were using the standard QED Lagrangian we would only have one diagram, where the interaction is mediated by the $\varepsilon$ mixing between the visible and dark field strengths. This leads to 
\begin{align}
\mathcal{M}_{JJ_D}=&\frac{e\, e_D\varepsilon q\,q_D}{k^2-m_D^2}J_D^\mu J_\mu~,\label{e.ampJJD}
\end{align}
We see that the final amplitude is that of a massive photon and that $J^\mu$ has the effective dark coupling of $e\varepsilon q/e_D$, just as expected. In the Zwanziger language the calculation is slightly more involved. As shown in Fig.~\ref{f.JJmix} there are two diagrams that contribute, one in which the mixing is through the $A$ fields and one in which the mixing is through the $B$ fields. The mixing terms in Eq.~\eqref{mixing} shows that these mixings are the same in form, for instance for the $A$-$A_D$ mixing we have
\beq
A_\alpha\frac{\varepsilon}{n^2}\left[ g^{\alpha\beta}\left(n\cdot k \right)^2+n^\alpha n^\beta k^2-\left(n\cdot k \right)\left(n^\alpha k^\beta+n^\beta k^\alpha \right)\right]A_{D\beta},
\eeq
but differing by a sign. However, the combination of these two diagrams leads to exactly the same result as Eq.~\eqref{e.ampJJD}:
\begin{align}
\mathcal{M}_{JJ_D}=&
\frac{e\, e_D\varepsilon q\,q_D}{n^2k^2(k^2-m_D^2)}\left[J_D^\mu J_\mu\left(n\cdot k\right)^2+k^2\left(n\cdot J\right)\left(n\cdot J_D\right) \right]\nonumber\\
&+\frac{e\, e_D\varepsilon q\,q_D}{n^2k^2(k^2-m_D^2)}\left\{J_D^\mu J_\mu\left[n^2k^2-\left(n\cdot k\right)^2\right]-k^2\left(n\cdot J\right)\left(n\cdot J_D\right) \right\}\nonumber\\
=&\frac{e\, e_D\varepsilon q\,q_D}{k^2-m_D^2}J_D^\mu J_\mu~,
\end{align}
where in the first (second) line we have written the results from the left (right) diagram of Fig.~\ref{f.JJmix}. Clearly, there is a nontrivial cancellation between the two diagrams which leads to the final result. We note that this has the expected decoupling properties, that is, in the limit $m_D\to \infty$ the amplitude vanishes. 

A similar calculation applies for the interaction between visible and dark magnetic charges. The diagrams in Fig.~\ref{f.JJmix} can still be used, but with $J\to K$ and $A\leftrightarrow B$. In this case we find the amplitude
\begin{align}
\mathcal{M}_{KK_D}=&-\frac{16\pi^2\varepsilon g\,g_D}{e\,e_Dn^2k^2(k^2-m_D^2)}\left\{K_D^\mu K_\mu\left[\left(n\cdot k\right)^2-n^2m_D^2\right]+k^2\left(n\cdot K\right)\left(n\cdot K_D\right) \right\}  \nonumber\\
&-\frac{16\pi^2\varepsilon g\,g_D}{e\,e_Dn^2k^2(k^2-m_D^2)}\left\{K_D^\mu K_\mu\left[n^2k^2-\left(n\cdot k\right)^2\right]-k^2\left(n\cdot K\right)\left(n\cdot K_D\right) \right\}\nonumber\\
=&-\frac{16\pi^2\varepsilon g\,g_D}{e\,e_Dk^2}K_D^\mu K_\mu~,\label{e.ampKKD}
\end{align}
where again the first (second) line is the result from the amplitude corresponding to the left (right) diagram of Fig.~\ref{f.JJmix}. Note that in this case the propagation is through a massless boson and that the effective coupling of the photon to the dark monopole is $-e\varepsilon g_D/e_D$ as expected from Eq.~\eqref{e.DiagCurrents}. While this agrees with the diagonalization analysis it does not make clear how the decoupling is manifest. Is it true that the dark photon can be removed from the spectrum leaving behind a magnetic coupling that violates the charge quantization condition? The answer is no, but the reason is a bit subtle, incorporating the bound nature of the magnetic charges. When $U(1)_D$ is broken by an electric mass term, like Eq.~\eqref{e.Emass}, the dark magnetic charges are confined, connected by tubes of dark magnetic flux. The tension in these flux tubes scales like $m_D^2$, and so in the limit $m_D\to\infty$ the magnetic charges are infinitely tightly bound. In such a limit the opposite charges of a monopoles and anti-monopole will cancel in every physical process. Thus, while the visible photon has a nonzero coupling to each constituent of the bound state, these couplings are opposite, and the state is point-like, so the bound state has zero coupling to the visible photon. This is an example of how the confined, bound state nature of the magnetic charges resolves confusions related to perturbative electric-magnetic interactions, which is the recurring theme of this work.

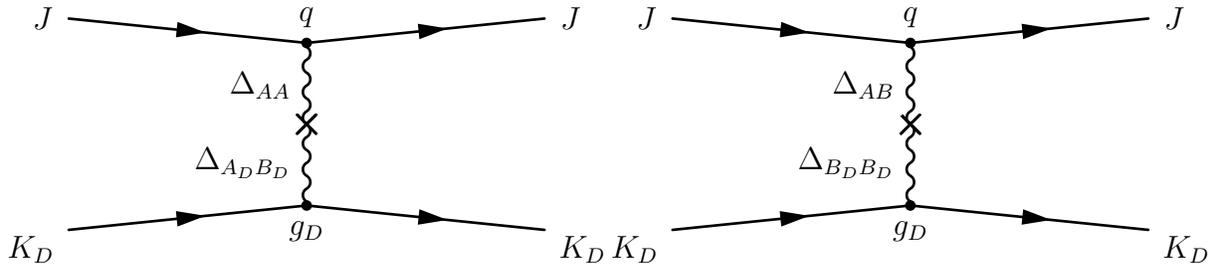
\begin{figure}[ht]
\centering
\begin{fmffile}{JKmix1}
\begin{fmfgraph*}(180,80)
\fmfpen{1.0}
\fmfstraight
\fmfleft{i1,p1,i2}\fmfright{o1,p2,o2}
\fmfv{l= $K_D$}{i1}\fmfv{l=$K_D$}{o1}
\fmfv{l= $J$,l.a=180}{i2}\fmfv{l=$J$,l.a=0}{o2}
\fmf{fermion,tension=1}{i1,v1,o1}
\fmf{fermion,tension=1}{i2,v2,o2}
\fmf{boson, tension=0.6,label=$\Delta_{A_DB_D}$,label.side=left}{v1,v}
\fmf{boson, tension=0.6,label=$\Delta_{AA}$,label.side=left}{v,v2}
\fmfv{decor.shape=circle,decor.filled=full,decor.size=1.5thick,l=$g_D$,l.a=-90}{v1}
\fmfv{decor.shape=circle,decor.filled=full,decor.size=1.5thick,l=$q$,l.a=90}{v2}
\fmfv{decor.shape=cross,decor.size=10,l.a=-25}{v}
\end{fmfgraph*}
\end{fmffile}\hspace{1.15cm}
\begin{fmffile}{JKmix2}
\begin{fmfgraph*}(180,80)
\fmfpen{1.0}
\fmfstraight
\fmfleft{i1,p1,i2}\fmfright{o1,p2,o2}
\fmfv{l= $K_D$}{i1}\fmfv{l=$K_D$}{o1}
\fmfv{l= $J$,l.a=180}{i2}\fmfv{l=$J$,l.a=0}{o2}
\fmf{fermion,tension=1}{i1,v1,o1}
\fmf{fermion,tension=1}{i2,v2,o2}
\fmf{boson, tension=0.6,label=$\Delta_{B_DB_D}$,label.side=left}{v1,v}
\fmf{boson, tension=0.6,label=$\Delta_{AB}$,label.side=left}{v,v2}
\fmfv{decor.shape=circle,decor.filled=full,decor.size=1.5thick,l=$g_D$,l.a=-90}{v1}
\fmfv{decor.shape=circle,decor.filled=full,decor.size=1.5thick,l=$q$,l.a=90}{v2}
\fmfv{decor.shape=cross,decor.size=10,l.a=-25}{v}
\end{fmfgraph*}
\end{fmffile}
\caption{\label{f.JKmix} The two diagrams describing the interaction between visible electric currents and dark magnetic currents. Scattering proceeds through $A$-$A_D$ mixing (left) and $B$-$B_D$ mixing (right).}
\end{figure}

We are now ready to consider the interaction between a visible electric charge and a dark magnetic charge. Here again there are two diagrams, see Fig.~\ref{f.JKmix}, one for each of $A$-$A_D$ and $B$-$B_D$ mixing terms. We find
\beq
\mathcal{M}_{JK_D}=4\pi q\frac{e\varepsilon }{e_D}g_D\frac{m_D^2}{k^2(k^2-m_D^2)}J^\mu K_D^\nu\frac{\varepsilon_{\mu\nu\alpha\beta} n^\alpha k^\beta}{n\cdot k}~,\label{e.ampJKD}
\eeq
which exhibits poles from both the massless and massive photons. This can also be understood from the diagonal basis, where there are also two diagrams, see Fig.~\ref{f.tchanSingle}. One involves the massless visible photon which has a small coupling to the dark monopoles while the other uses the small coupling of the dark photon to visible electric charges. In the decoupling limit, we are left with only the visible photon interaction, but as before the bound nature of the monopoles conspires to eliminate the photon's coupling to the point-like bound state.

\section{Discrete Symmetries and Magnetic Charges\label{s.Symmetries}}
Before proceeding we describe the action of the discrete symmetries parity, time reversal and charge conjugation: $P$, $T$, and $C$. Ramsey noted~\cite{Ramsey:1958gvj}, and Weinberg clarified \cite{Weinberg:1965rz}, that consistency with the $CPT$ theorem and the Maxwell equations (\ref{e.MaxEq}) requires that both $P$ and $T$ must transform magnetically charged particles into their antiparticles. This observation motivates using two separate charge conjugation operators: $C_E$ which takes electric charges to their antiparticles but acts trivially on magnetic particles and $C_M$ which conjugates magnetic charges but acts trivially on electric particles. We also define $P_E$ and $T_E$ as the conventional parity and time reversal operators which are merely the conventional space-time transformations. 

The total charge conjugation operator, which takes electric \emph{and} magnetic particles to their antiparticles is simply $C=C_EC_M$. We also define the total parity and time-reversal operators as $P=C_MP_E$ and $T=C_MT_E$, respectively. These operators act on electric charges in the usual way, but when acting upon magnetic particles they include the necessary conjugation of the magnetic charges. This formulation allows one to easily find the  behavior of magnetic particles, currents, and bounds states under discrete symmetries directly from the familiar electric case. 

Now consider the discrete transformation properties of all the fields and currents.
The $P$ and $T$ transformations of the electric and magnetic fields, $E^i$ and $H^i$, follow from the Lorentz transformations of $F^{\mu\nu}$. Both the electric and the magnetic  currents, $J^\mu$ and $K^\mu$, are conserved currents, as seen from Maxwell's equations~\eqref{e.MaxEq}. This forbids using an axial current for $K^\mu$, as has been tried in the past \cite{Taylor:1967fsz,Acharya:1973un}. Therefore, both currents must be proportional to the vector currents formed from the correspondingly charged fields. Because charge conjugation also takes particles to their antiparticles, we see that $E^i$ and $H^i$ must be odd under $C$. 
The various transformations are summarized in Table \ref{PCT}.

\begin{table}[htb]
	\begin{center}
		\begin{tabular}{c|ccccc}
		 &  $C$ & $CP$ & $P$ & $T$    \\ \hline
		 $E^i$ &$-1$ & $+1$  & $-1$ & $+1$   \\
		 $H^i$ &$-1$ &  $-1$  &  $+1$ & $-1$    \\
		 $A^0$ &$-1$ &  $-1$  & $+1$  &$+1$    \\
		 $B^0$ &$-1$ &  $+1$  & $-1$ & $-1$     \\
		 $A^i$  &$-1$ &  $+1$  & $-1$ & $-1$    \\
		 $B^i$  &$-1$ &  $-1$  & $+1$ & $+1$    \\
		 $J^0$ &$-1$ &  $-1$ & $+1$  & $+1$     \\
		$K^0$ &$-1$ &  $+1$  & $-1$ & $-1$     \\
		 $J^i$  &$-1$ &  $+1$ & $-1$ & $-1$    \\
		 $K^i$  &$-1$ &  $-1$ & $+1$ & $+1$  
		\end{tabular}
	\caption{Discrete symmetry transformations of electromagnetic fields, potentials, and currents.
	\label{PCT}}
	\end{center}
\end{table}

The currents must be constructed so as to satisfy the above requirements explicitly. The electric scalar current is defined in terms of the covariant derivative $D_\mu$ as
\beq
J^\mu_S=iq\left[\phi^\dagger D_\mu\phi-\phi\left( D_\mu\phi\right)^\dagger \right],
\eeq
and of course $CP$ conjugation is given by $\phi({\bf x},t) \xrightarrow{CP}\phi^\dagger(-{\bf x},t)$. Similarly, under parity we have 
$\phi({\bf x},t) \xrightarrow{P}\eta^*\, \phi^\dagger(-{\bf x},t)$.
Similarly, for fermions we have
\beq
J_F^\mu=q\overline{\psi}\gamma^\mu\psi,
\label{electriccurrent}
\eeq
where $q$ is the electric charge of the fermion. The transformation properties are well known from any introductory text on quantum field theory. In either case, the currents transform exactly as shown in Table~\ref{PCT}.
It then follows that if 
\beq
A_0\xrightarrow{CP} -A_0, \ \ A_i\xrightarrow{CP} A_i~,
\eeq
then the interaction $A\cdot J$ is a $CP$ invariant. 

We now turn to the magnetic currents. From Maxwell's equations~\eqref{e.MaxEq}, and from the fact that ${}^\ast\! F^{\mu\nu}$ has opposite $CP$ to $F^{\mu\nu}$ (because of the Levi-Civita tensor) we see that $K^\mu$ must have the opposite $CP$ to $J^\mu$. Consequently the magnetic gauge potential, $B_\mu$ must also have the opposite $CP$ to $A_\mu$:
\beq
B_0\xrightarrow{CP} B_0, \ \ B_i\xrightarrow{CP} -B_i~,
\eeq 
in agreement with Weinberg's more general analysis \cite{Weinberg:1965rz}.
This is also required by the $A$-$B$ mixing term in the Lagrangian (\ref{e.zLag}).
But how can conserved magnetic currents have the correct transformation laws? Consider the current for a magnetically charged scalar field $\phi_m$
\beq
K^\mu_S=ig\,\left[\phi_m^\dagger D^\mu\phi_m-\phi_m\left(D^\mu\phi_m\right)^\dagger \right]~.
\eeq
This naively seems to have exactly the same $CP$ behavior as $J^\mu_S$. If, however, $CP$ acts on magnetically charged fields as
\beq
\phi_m({\bf x},t)\xrightarrow{CP} \phi_m(-{\bf x},t), \ \ \phi_m^\dagger({\bf x},t)\xrightarrow{CP} \phi_m^\dagger(-{\bf x},t)~,
\eeq 
then $K^\mu$ has the correct transformation properties. This implies that parity takes a magnetic particle to its antiparticle
\beq
\phi_m({\bf x},t)\xrightarrow{P} \phi_m^\dagger(-{\bf x},t), \ \ \phi_m^\dagger({\bf x},t)\xrightarrow{P} \phi_m(-{\bf x},t)~,
\eeq 
as required by Weinberg \cite{Weinberg:1965rz}. This is also in agreement with taking $P=C_MP_E$ and $CP=C_EP_E$, as anticipated earlier.

For magnetically charged fermions $\psi_m$ the conserved fermionic current is
\beq
K^\mu=g\,\overline{\psi}_m\gamma^\mu\psi_m~.
\eeq
To match the behavior of $B_\mu$ we again need to effectively exchange the behavior of electric particles under $P$ with $CP$. This is simply accomplished by acting with $P=C_MP_E$ and $CP=C_EP_E$ on this current, which leads precisely to the required transformation properties outlined in Table~\ref{PCT}.

 Though we do not pursue such theories in this paper, we note that if a theory contains a field with both electric and magnetic charges (a dyon) then $CP$ invariance puts additional restrictions on the spectrum \cite{Witten,Csaki:2010rv}. In addition, if a $\theta$ term is included in the dynamics, which induces an electric charge \cite{Witten}  for a monopole proportional to $\theta  g$, then $CP$ is explicitly broken for generic values of $\theta$.

Our eventual discussion of magnetic pair production relies in part upon understanding the discrete symmetries of particle--antiparticle bound states.
Since the electric and magnetic currents transform differently, magnetic bound states are not simply relabelings of electric bound states. 
In bound states of an electrically charged particle and antiparticle the $P$ and $C$ eigenvalues\footnote{This assumes a single flavor of particle, if the bound state includes different flavors then the discussion needs to be extended to include something like $G$-parity, as is done for the mesons of QCD.}  are given by
\begin{align}
P&=(-1)^{L+1},  &C=&(-1)^{L+S}, &&\text{Electric Fermion Bound State},\\
P&=(-1)^{L},  &C=&(-1)^{L}, &&\text{Electric Scalar Bound State},
\end{align}
where $L$ is the orbital angular momentum of the state. The parity of the fermion--antifermion bound state reflects the fact that fermions and antifermions have opposite intrinsic parity.

Bound states of magnetically charged particles and antiparticles behave differently. These differences are  most simply understood by considering $P=C_MP_E$ and $C=C_MC_E$ as outlined above. We see that the results for charge conjugation carry through, however, parity is quite different:
\begin{align}
P&=(-1)^{S+1}, &C=&(-1)^{L+S}, &&\text{Magnetic Fermion Bound State},\\
P&=+1,  &C=&(-1)^{L}, &&\text{Magnetic Scalar Bound State}.
\end{align}
Of course, one can also derive the same results directly from the definitions of the fields in terms of annihilation and creation operators. A summary for low angular momentum states is given in Table~\ref{dLJ}.

\begin{table}[ht]
	\begin{center}
		\begin{tabular}{c|cc}
		${}^d\! L_J$ & \multicolumn{2}{c}{Fermionic $J^{PC}$} \\
		 &electric &magnetic  \\ \hline 
		 ${}^1\! S_0$ &$0^{-+}$&$0^{-+}$ \\
		 ${}^3\! S_1$ &$1^{--}$&$1^{+-}$ \\
		 ${}^1\! P_1$ &$1^{ + -}$&$1^{--}$ \\
		 ${}^3\! P_0$ &$0^{+ +}$&$0^{++}$ \\
		 ${}^3\! P_1$ &$1^{++}$&$1^{++}$ \\
		  ${}^3\! P_2$ &$2^{++}$&$2^{++}$ \\
		 ${}^3\! D_1$ &$1^{--}$&$1^{+-}$ \\
		\end{tabular}\;\;\;
		\begin{tabular}{c|cc}
		${}^d\! L_J$ & \multicolumn{2}{c}{Scalar $J^{PC}$} \\
		 &electric &magnetic  \\ \hline 
		 ${}^1\! S_0$ &$0^{++}$&$0^{++}$ \\
		 ${}^1\! P_1$ &$1^{ - -}$&$1^{+-}$ \\
		 ${}^1\! D_2$ &$2^{++}$&$2^{++}$\vspace{2.025cm}\\
		\end{tabular}
	\end{center}
	\caption{$J^{PC}$ quantum numbers for bound states of electric or magnetic particles with bound state spin degeneracy $d$, orbital angular momentum state $L$, and total angular momentum $J$. Bound states composed of fermions (scalars) are on the left (right).\label{dLJ}}
\end{table}

\section{Electric-Magnetic Scattering\label{s.scat}}
As shown in previously, when kinetically mixed to a dark photon with an electric mass, the visible photon couples to confined magnetic monopoles of the dark $U(1)_D$. These bound states are composed of a monopole--anti-monopole pair connected by a tube of dark magnetic flux. We now consider perturbative $t$-channel, elastic scattering of a visible sector electric charge with the dark magnetic bound state. 

If one assumed factorization and tried to calculate the amplitude for  a particle with electric charge $q$, scattering off a single magnetic monopole in the bound state, with charge $g$ one finds (using the mixed charge propagator in Eq.~\eqref{e.mixedProp}) for fermions and scalars:
\begin{align}
\mathcal{M}_F=&qg\overline{u}(p^{\prime}_f)\gamma^\mu u(p^{\prime}_i)\frac{\varepsilon_{\mu\nu\alpha\beta}n^\alpha k^\beta}{k^2(n\cdot  k)}\overline{u}(p_f)\gamma^\nu u(p_i),\label{e.tChanSingle}\\
\mathcal{M}_S=&qg\left( p'_f+p'_i\right)^\mu\frac{\varepsilon_{\mu\nu\alpha\beta}n^\alpha k^\beta}{k^2(n\cdot  k)}\left( p_f+p_i\right)^\nu,\label{e.tChanScalSingle}
\end{align}
where $k^\mu$ is the momentum transfer while $p^\mu_{i,f}$ and $p^{\prime \mu}_{i,f}$ are the initial and final momenta of the electric and magnetic particles, respectively. As can be checked by direct computation, the spurious pole at $n\cdot k $ does not cancel in the squared amplitude \cite{Weinberg:1965rz}. The appearance of unphysical poles in the amplitude suggests that factorization cannot hold.

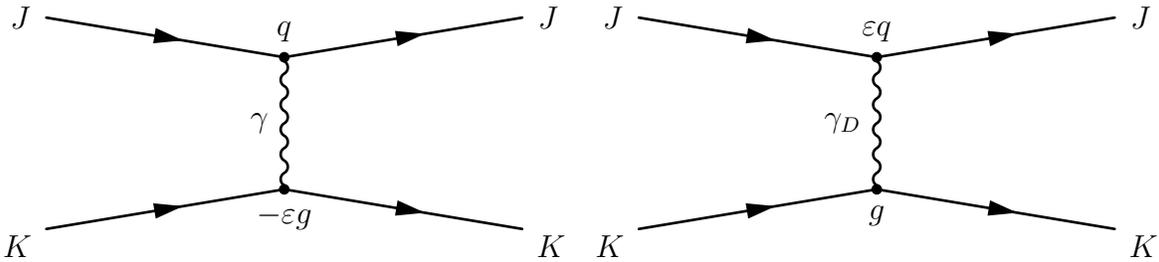
\begin{figure}[ht]
\centering
\begin{fmffile}{EMtchan3}
\begin{fmfgraph*}(180,80)
\fmfpen{1.0}
\fmfstraight
\fmfleft{i1,p1,i2}\fmfright{o1,p2,o2}
\fmfv{l= $K$}{i1}\fmfv{l=$K$}{o1}
\fmfv{l= $J$,l.a=180}{i2}\fmfv{l=$J$,l.a=0}{o2}
\fmf{fermion,tension=1}{i1,v1,o1}
\fmf{fermion,tension=1}{i2,v2,o2}
\fmf{boson, tension=0.6,label=$\gamma$,label.side=left}{v1,v2}
\fmfv{decor.shape=circle,decor.filled=full,decor.size=1.5thick,l=$-\varepsilon g$,l.a=-90}{v1}
\fmfv{decor.shape=circle,decor.filled=full,decor.size=1.5thick,l=$q$,l.a=90}{v2}
\end{fmfgraph*}
\end{fmffile}\hspace{1cm}
\begin{fmffile}{EMtchan4}
\begin{fmfgraph*}(180,80)
\fmfpen{1.0}
\fmfstraight
\fmfleft{i1,p1,i2}\fmfright{o1,p2,o2}
\fmfv{l= $K$}{i1}\fmfv{l=$K$}{o1}
\fmfv{l= $J$,l.a=180}{i2}\fmfv{l=$J$,l.a=0}{o2}
\fmf{fermion,tension=1}{i1,v1,o1}
\fmf{fermion,tension=1}{i2,v2,o2}
\fmfv{decor.shape=circle,decor.filled=full,decor.size=1.5thick,l=$g$,l.a=-90}{v1}
\fmfv{decor.shape=circle,decor.filled=full,decor.size=1.5thick,l=$\varepsilon q$,l.a=90}{v2}
\fmf{boson,label=$\gamma_D$,label.side=left,tension=0.6}{v1,v2}
\end{fmfgraph*}
\end{fmffile}
\caption{\label{f.tchanSingle} The $t$-channel scattering of an electrically charged particle and a bound magnetic monopole. The two diagrams from the exchange of the visible photon and the dark photon give  amplitudes with  opposite sign.}
\end{figure}

In fact, the amplitudes in Eqs.~\eqref{e.tChanSingle} and \eqref{e.tChanScalSingle} only includes the exchange of the visible photon. One must also include a second diagram mediated by the massive dark photon, as seen in Fig.~\ref{f.tchanSingle}. As the diagonal currents given in Eq.~\eqref{e.DiagCurrents} make clear, the processes involving each of the photons are almost identical, except they enter with opposite sign from the $\varepsilon$ suppressed couplings and the dark photon has a mass. Thus, by including both photons the $1/k^2$ pole we expect from a single massless photon becomes, in the squared amplitude,
\beq
\frac{1}{k^4}\to \left(\frac{1}{k^2}-\frac{1}{k^2-m_D^2}\right)^2=\frac{m_D^4}{k^4\left(k^2-m_D^2 \right)^2}~,
\eeq
 as we might have expected from the mixed basis calculation in Eq.~\eqref{e.ampJKD}. Clearly, scattering is highly suppressed for momentum transfers much greater than the dark photon mass. This agrees with our intuition that it is only for momentum transfers below the mass of the dark photon that the visible photon has an effective coupling to the dark monopoles. While including the dark photon does nothing to eliminate the spurious pole in the amplitude, it does show that the scattering is dominated by longer wavelength photons which are sensitive to the whole of the bound state.

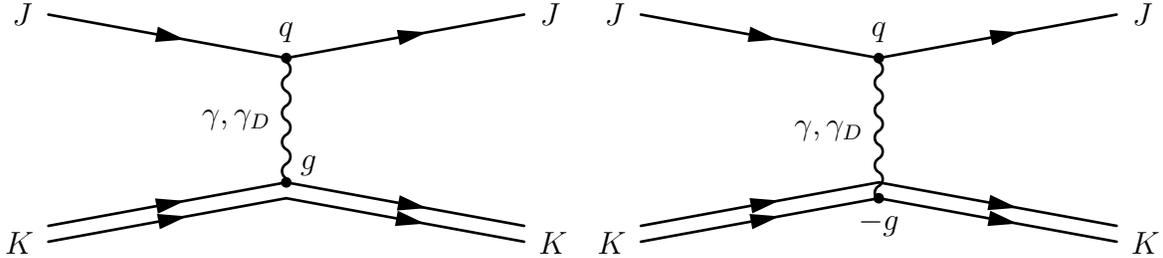
\begin{figure}[t]
\centering
\begin{fmffile}{EMtchan1}
\begin{fmfgraph*}(180,80)
\fmfpen{1.0}
\fmfstraight
\fmfleft{i1,p1,i2}\fmfright{o1,p2,o2}
\fmfv{l= $K$}{i1}\fmfv{l=$K$}{o1}
\fmfv{l= $J$,l.a=180}{i2}\fmfv{l=$J$,l.a=0}{o2}
\fmf{fermion,tension=1}{i1,v1,o1}
\fmf{fermion,tension=1}{i2,v2,o2}
\fmf{boson, tension=0.7,label=$\gamma,,\gamma_D$,label.side=left}{v1,v2}
\fmffreeze
\fmfi{fermion}{vpath (__i1,__v1) shifted (thick*(0,-3))}
\fmfi{fermion}{vpath (__v1,__o1) shifted (thick*(0,-3))}
\fmfv{decor.shape=circle,decor.filled=full,decor.size=1.5thick,l=$g$,l.a=25}{v1}
\fmfv{decor.shape=circle,decor.filled=full,decor.size=1.5thick,l=$q$,l.a=90}{v2}
\end{fmfgraph*}
\end{fmffile}\hspace{1cm}
\begin{fmffile}{EMtchan2}
\begin{fmfgraph*}(180,80)
\fmfpen{1.0}
\fmfstraight
\fmfleft{i1,p1,i2}\fmfright{o1,p2,o2}
\fmfv{l= $K$}{i1}\fmfv{l=$K$}{o1}
\fmfv{l= $J$,l.a=180}{i2}\fmfv{l=$J$,l.a=0}{o2}
\fmf{fermion,tension=1}{i1,v1,o1}
\fmf{fermion,tension=1}{i2,v2,o2}
\fmf{phantom, tension=0.7}{v1,v2}
\fmf{phantom,tension=0.7}{p1,v3,p2}
\fmffreeze
\fmfi{fermion}{vpath (__i1,__v1) shifted (thick*(0,-3))}
\fmfi{fermion}{vpath (__v1,__o1) shifted (thick*(0,-3))}
\fmfv{decor.shape=circle,decor.filled=full,decor.size=1.5thick,l=$q$,l.a=90}{v2}
\fmfiv{decor.shape=circle,decor.filled=full,decor.size=1.5thick,l=$-g$,l.a=-90,l.d=5}{vloc(__v1) shifted (thick*(0,-3))}
\fmfshift{(0,-30)}{v3}
\fmf{boson,label=$\gamma,,\gamma_D$,label.side=right,tension=0.7}{v2,v3}
\end{fmfgraph*}
\end{fmffile}
\caption{\label{f.tchan} The $t$-channel scattering of an electrically charged particle and a bound magnetic monopole and anti-monopole.  There are two diagrams with the exchange of the visible photon and two with the exchange of the dark photon.}
\end{figure}

To see how the spurious pole can cancel it is instructive to first consider the static limit, where the monopole and anti-monopole are infinitely heavy, and separated by a distance $L$ along the direction $\hat{n}$, see Fig.~\ref{f.tchan}. Combining the static limit with the low-energy limit for the momentum transfer (so that we can, temporarily, neglect the dark photon) reduces the problem to that of an electric charge scattering in the background field of a magnetic dipole. In this case the amplitude can be determined without using the self-dual formalism simply by finding the vector potential (in the standard one-potential formalism) produced by the monopoles.  The amplitude is given by
\beq
\mathcal{M}=eq\overline{u}(p_f)\slashed{A}(k)u(p_i)
\eeq
where $\vec{A}$ is the Fourier transform of the vector potential of the dipole field. In position space the vector potential  is \cite{Dirac,Jordan}
\beq
\vec{A}(\vec{r}\,)=\frac{g}{4\pi}\int_\text{string}\!\!d\vec{\ell}'\times\frac{\vec{r}-\vec{r}'}{|\vec{r}-\vec{r}'|^3},
\eeq
where the integral is taken along the string. For a straight string of length $L$ along the direction $\hat{n}$, the Fourier transform of this potential (after shifting the integration variable to $\vec{r}-{\vec{r}}^{\,\prime}$) is found to be
\beq
\vec{A}(\vec{k}\,)=\frac{g}{\vec{k}^2}\left(1-e^{-i L \hat{n}\cdot\vec{k}}\right)\frac{\hat{n}\times\vec{k}}{\hat{n}\cdot\vec{k}}~.\label{e.background}
\eeq
The two terms correspond to Dirac potentials for the monopole and anti-monopole.
Note that this expression is finite in the $\vec{k}\cdot\hat{n}\to0$ limit because of the exponential factor. This factor accounts for the relative phase between the interactions with the constituents due to propagating the extra distance $L$.

We can go beyond the low-energy limit by including the dark photon exchange diagrams in Fig.~\ref{f.tchan} while retaining the static limit and keeping track of the relative phase between the monopole and anti-monopole contributions by hand. Then, using the mixed propagator~\eqref{e.mixedProp} we find the sum of these diagrams is
\beq
\mathcal{M}=qg\left(1-e^{-i L n\cdot k}\right)\overline{u}(p^{\prime}_f)\gamma^\mu u(p^{\prime}_i)\frac{m_D^2\varepsilon_{\mu\nu\alpha\beta}n^\alpha k^\beta}{k^2(k^2-m_D^2)(n\cdot  k)}\overline{u}(p_f)\gamma^\nu u(p_i)~.
\eeq
The similarity to Eq.~\eqref{e.background} is obvious, as is the finite nature of the amplitude in the limit where $n\cdot k \to0$. 

So far we have taken the monopole masses to be infinite and kept them separated by a fixed length, but we are really interested in finite mass monopoles that are part of a dynamical bound state. Even though we are imagining that the  monopole and anti-monopole are different flavors---so that we have a stable bound state, as in ref. \cite{Terning:2019bhg}--- let us first consider the case where the masses are degenerate, and return to the general case near the end of this section. A magnetic dipole moment for such a quantum bound state can only occur when the bound state is polarized by an external magnetic field.  For simplicity we restrict ourselves to non-relativistic bound states; this means that the string tension $\sim m_D^2$ is much smaller than the monopole mass squared, and that the coupling of the dark photon to the monopole is perturbative.

In the Born approximation the non-relativistic amplitude for elastic scattering off a bound particle with wavefunction $\phi(x^\prime)$ is given by
\beq
{\mathcal M}&=& \int d^3x\, d^3x^\prime e^{-i {\vec p_f}\cdot {\vec x}}\phi^*(x^\prime)  \Delta(x-x^\prime)  e^{i {\vec p_i}\cdot {\vec x}}\phi (x^\prime), \\
&=& \int d^3x\, e^{-i {\vec k}\cdot {\vec x}} \int d^3x^\prime   \Delta(x-x^\prime)  |\phi (x^\prime)|^2,
\eeq
where the momentum transfer is $k^\mu=p_f^\mu-p_i^\mu$,  the scattering particle's initial and final wavefunctions are  free plane waves, and $\Delta(x-x')$ is the propagator of the interaction between free and bound particles. For an electron scattering off a single bound electron we have
\beq
\Delta(x-x')= \frac{\alpha}{|x-x'|}~,
\eeq
where $\alpha$ is the fine structure constant, since in the non-relativistic limit the current is approximated by the charge density.
In the case of interest here, $\Delta(x-x')$ is the Fourier transform of  the mixed charge propagator~\eqref{e.mixedProp}, but we can't take both vertices to be charge densities, because of the Levi-Civita tensor than connects the two currents. However, in the non-relativistic limit, with a monopole much heavier than the electron, we can approximate the scalar amplitude in Eq.~\eqref{e.tChanScalSingle} using
\beq
K^\mu=g(p'_f+p'_i)^\mu\approx 2g(M,0,0,0),
\eeq
while the electric current of the scattering charged particle is
\beq
J^\mu=q(p_f+p_i)^\mu \approx q(2m,\vec{v}_f+\vec{v}_i),
\eeq
to leading order in the relative velocity of the electric and magnetic particles. So, the interaction is proportional to $qgM(\vec{v}_f+\vec{v}_i)\cdot \vec{n}\times\vec{k}$, where the velocity dependance is as expected from undergraduate E\&M. It is related to the $\vec{v}\times\vec{H}$ part of the familiar Lorentz force law, where $\vec{H}$ is the magnetic field. Thus, in the electric-magnetic scattering case we are still sensitive to both charge densities, but also to their relative velocities. Of course, in the opposite regime, where the electric charge is heavy and the magnetic particles are traveling with some small velocity there is a similar, velocity suppressed effect. This agrees with the, less familiar, $\vec{v}\times\vec{E}$ term in the Lorentz force law for a magnetic charge. In what follows we simply define $\Delta(x-x')$ to include this velocity dependence when appropriate.

In the standard electric-electric case, changing variables to $y=x-x^\prime$ we have
\beq
{\mathcal M}&=&  \int d^3y \,e^{-i \vec{k}\cdot \vec{y}}  \Delta(y) \int d^3x^\prime  e^{-i \vec{k}\cdot \vec{x}^\prime}  |\phi (x^\prime)|^2\nonumber\\
&=&\int d^3y \,e^{-i \vec{k}\cdot \vec{y}}  \Delta(y) \,F(k)
\eeq
which is the fixed target scattering amplitude multiplied by a form factor, $F(k)$. 

Generalizing to two bound particles (a particle and an antiparticle with opposite charge) the wavefunction now depends on two positions, $\phi(x_p,x_{\overline p})$. We can choose center of mass coordinates:
\beq
0=m_p\, x_p+m_{\overline p}\, x_{\overline p} ~,\quad\quad x\equiv x_p-x_{\overline p}~,\\
x_p= \frac{m_{\overline p}}{m_p+m_{\overline p}}\,x~,\quad\quad x_{\overline p}=- \frac{m_p}{m_p+m_{\overline p}}\,x~,\label{e.Xp}
\eeq
where $x$ is  the relative coordinate between the two particle positions.
This reduces the problem to a one particle Schr\"{o}dinger equation  with a reduced mass
\beq
\mu=\frac{m_p\,m_{\overline p}}{m_p+m_{\overline p}}~,
\eeq
and a wavefunction $\phi(x)$. It is convenient to introduce particle and antiparticle charge densities. Since $\phi(x)$ is normalized:
\beq
\int d^3x\,|\phi(x)|^2 =1~,
\eeq
and we want the charge densities to satisfy
\beq
g\int d^3x\,|\phi(x)|^2 =\int d^3x_p\,\rho_{p}(x_p)=g  ~,\quad\quad -g\int d^3x\,|\phi(x)|^2 =\int d^3x_{\overline p}\,\rho_{\overline p}(x_{\overline p})=-g ~,
\eeq
we find
\begin{align}
\rho_{p}(x_p)=&g\left( \frac{m_p+m_{\overline p}}{m_{\overline p}}\right)^3  \left|\phi\left( \frac{m_p+m_{\overline p}}{m_{\overline p}}\, x_p\right)\right|^2~,\\
\,\rho_{\overline p}(x_{\overline p})=&-g \left( \frac{m_p+m_{\overline p}}{m_p}\right)^3 \left|\phi\left(- \frac{m_p+m_{\overline p}}{m_p}\,x_{\overline p}\right)\right|^2.
\end{align}

For the time being we restrict our discussion to the degenerate case $m_p=m_{\overline p}$.
We then combine the contributions from each charge density:
\begin{align}
{\mathcal M}&=\int d^3y \,e^{-i \vec{k}\cdot \vec{y}}\left[ \int d^3x_p  e^{-i \vec{k}\cdot \vec{x}_p}  \Delta(y) \rho_p (x_p)+\int d^3x_{\overline{p}}  e^{-i \vec{k}\cdot \vec{x}_{\overline{p}}}  \Delta(y) \rho_{\overline{p}} (x_{\overline{p}})   \right]\nonumber\\
&=\int d^3y \,e^{-i \vec{k}\cdot \vec{y}} \int d^3x^\prime  e^{-i \vec{k}\cdot \vec{x}^\prime}  \Delta(y) \left[\,  \rho_p (x^\prime) +\rho_{\overline p} (x^\prime)\right]~.
\end{align}
In the electric-electric case we would write
\beq
{\mathcal M}&=&\int d^3y \,e^{-i \vec{k}\cdot \vec{y}}  \Delta(y) \,F_{p{\overline p}}(k)= \Delta(k) F_{p{\overline p}}(k)~,
\label{Formfactordef}
\eeq
where $\Delta(k) $ is Fourier transform of $\Delta(y) $.
However for electric-magnetic scattering we need to be more careful with the ${\vec n}$ dependence.
To focus on this new effect, we separate this dependance from the rest of the propagator as
\beq
\Delta(k) ={\widetilde \Delta}(k) \cdot \frac{ {\vec n} \times {\vec k} }{ {\vec n} \cdot {\vec k}}~.
\eeq
This leads to 
\beq
{\mathcal M}&=& {\widetilde \Delta}(k) \cdot \int d^3x\,  e^{-i \vec{k}\cdot \vec{x}} \frac{ {\vec n} \times {\vec k} }{ {\vec n} \cdot {\vec k}} \left[\,  \rho_p (x) + \rho_{\overline p} (x)\right]~,
\label{Formfactormag}
\eeq
We have kept $\vec{n}$ inside the integral because our association of $\vec{n}$ with the direction along the flux tube connecting the monopoles  implies
\beq
\vec{n}\propto \vec{x}~.
\eeq
From Eq.~\eqref{e.Xp} we see this means that $\vec{n}\propto \vec{x}_{p,\overline{p}}$, so we can rewrite the form factor as
\beq
F_{p{\overline p}}(k)= \int d^3x  e^{-i \vec{k}\cdot \vec{x}} \frac{ {\vec x}\times {\vec k} }{ {\vec x} \cdot {\vec k}} \left[\,  \rho_p (x) + \rho_{\overline p} (x)\right]
\eeq
Assuming parity is a good symmetry means that the magnetic particle and antiparticle have the same spatial distribution (remember that parity acts like $CP$ for electric particles) so 
\beq
\rho_p(x) = -\rho_{\overline p}(x),
\eeq
which implies that
\beq
F_{p{\overline p}}(k)=0~.
\eeq
Thus, when the bound state is in a configuration which is symmetric under the interchange of its constituents, the scatterer feels as much of the positive charge as the negative one, and there is no contribution to the scattering at this order. No specific cancellation of the spurious pole is needed as the whole amplitude vanishes.

In the presence of  an external magnetic field, however, the energy eigenstates will have dipole moments  aligned with the field. In other words, in this configuration the bound monopoles have a non-zero dipole moment, since $\rho_p(x) \ne -\rho_{\overline p}(x)$. The dipole moment is given by
\beq
g L \langle{\hat n} \rangle=\int d^3x \, {\vec x}\left[\,  \rho_p (x) + \rho_{\overline p} (x)\right]~,\label{e.dipole}
\eeq
where $L$ is the expectation value of the distance between the two charges and $\langle \hat{n}\rangle$ is the expectation value of the vector pointing from the negative charge to the positive. Note that since the string is effectively oriented along the magnetic field we must choose Zwanziger's spacelike vector ${\hat n}$ along this string direction. When the size of the bound state is much smaller than the wavelength of the photon, $1/|{\vec k}|$, we can Taylor expand the exponential in the form factor:
\beq
F_{p{\overline p}}(k)=\int d^3x  e^{-i\vec{k}\cdot \vec{x}} \, \frac{ {\vec x}' \times {\vec k} }{ {\vec x}' \cdot {\vec k}} \left[\,  \rho_p (x) +  \rho_{\overline p} (x)\right]~.
\eeq
Since the particles are identical except for the sign of the charge, parity  implies
\beq
\int d^3x' \,\rho_p(x') = - \int d^3x' \,\rho_{\overline p}(x')~.
\eeq
In other words, we have that $\rho_p (x) +  \rho_{\overline p} (x) $ and hence
\beq
\frac{ {\vec x} \times {\vec k} }{ {\vec x} \cdot {\vec k}} \left[\rho_p (x) +  \rho_{\overline p} (x)\right]~,
\eeq
is an odd function of $x$. Therefore, the first term in the series vanishes, as well as all even powers of $\vec{x}$ in the expansion of the exponential. Since every remaining term has at least one power of $\vec{x}\cdot\vec{k}$ we see that the spurious pole is cancelled. We also find that the leading effect is proportional to the dipole term
\beq
F_{p{\overline p}}(k) \approx -i\int d^3x\,  {\vec k}\times {\vec x} \left[\, \rho_p (x) +  \rho_{\overline p} (x)\right] =i gL\langle {\hat n}\rangle \times {\vec k}~,
\eeq
which matches the static dipole result in Eq.~\eqref{e.background} when we expand to the same order in $k$. While higher order terms may be more complicated, we have already seen that the vanishing of the leading term, i.e. $F_{p{\overline p}}(0)=0$ as expected for charge neutral states, is enough to guarantee that the spurious pole cancels.

Higher-order odd powers in the expansion will also include a factor of $\langle {\hat n}\rangle \times {\vec k}$. This can be seen by considering the tensor structure of each term in the expansion. Using the usual index notation we can write the $N$th term in the expansion as proportional to
\beq
k_j\epsilon^{jkl}k_{i_1}\ldots k_{i_N}\int dx\, x_{i_1}\ldots x_{i_N} x_k\left[\,  \rho_p (x) +  \rho_{\overline p} (x)\right]=k_j\epsilon^{jkl}k_{i_1}\ldots k_{i_N} f_{i_1\cdots i_N,k}~,
\eeq
where the tensor structure of the unknown function $f_{i_1\cdots i_N,k}$ can only be products of $\langle n_i\rangle$ and $\delta_{ij}$.\footnote{The product is completely symmetric under exchange of $x_i$, so the Levi-Civita tensor cannot appear.} Since all the non-vanishing contributions have $N$ odd there is always a factor of $\langle n_k\rangle$ which leads to one factor of $\langle\hat{n}\rangle\times \vec{k}$ after integration. Thus, there is no pole as ${\hat n} \cdot {\vec k}\to 0$, while there can still be finite effects from higher multipole moments. In addition, we find that when $\langle\hat{n}\rangle\times \vec{k}=0$ there is no scattering.

We now consider the case where the masses are not degenerate.
At very low momentum transfer the separation of the charges cannot be resolved, so we still have $F_{p{\overline p}}(0) =0$. As we have seen in the previous cases, this is enough to cancel the spurious pole inside the integral. Because of the asymmetry in the masses, however, one would expect non-zero scattering without an external magnetic field. One might expect the bound state to have a magnetic charge radius. Let's consider the  ground state in some detail by making use of its rotational symmetry. Choosing coordinates with the $z$-axis along $k$ we have the first term in the Taylor series of the exponential:
\beq
F^{(1)}_{p{\overline p}}(k)=-i \int r^2 dr d\Omega\,  k r \cos \theta \, \frac{ k \sin \theta (\sin\varphi,-\cos \varphi,0) }{ k \cos \theta} \left[\,  \rho_p (x) +  \rho_{\overline p} (x)\right]~,
\eeq
which vanishes when the $\varphi$ integration is performed. In fact, the $\varphi$ integration causes the form factor to vanish for all powers of $r$ in the expansion. This implies that the form factor vanishes completely for spherically (or even just axially) 
 symmetric bound states, and there is no magnetic charge radius when probed by electric charges. 

Finally, removing the restriction of elastic scattering we can have transitions between different confined states even in the absence of an external magnetic field. 
In order for the transition to occur there must be a transition dipole moment (at leading order). To fully calculate this process one would need to go beyond the Zwanziger self-dual formalism and incorporate a dynamical string \cite{Lechner:1999ga}.
It seems plausible, however, that using the formalism above with ${\hat n}$ aligned parallel to the transition dipole moment would give a good approximation to the amplitude. Again for small ${\hat n}\cdot{\vec k}$ one can Taylor expand the exponential and this is sufficient to cancel the spurious pole and lead to finite predictions for inelastic scattering.


\section{Monopole Pair Production\label{s.Prod}}
The simplest process for monopole production is the annihilation of two electrically charged particles into an off-shell photon that then produces a monopole--anti-monopole pair. This is not simply a crossing ($t$-channel to $s$-channel) of the scattering considered in Fig.~\ref{f.tchan}, since the $t$-channel scattering included monopole--anti-monopole pairs in both the initial and final state. Thus, we need to treat production separately. In fact, while for scattering we need to invoke confinement to ensure that the total magnetic charge of a physical state is zero, for pair production charge conservation itself is enough to ensure that the total charge of the produced particles is zero. Thus, we can discuss this case without any requirements on details of the bound state wavefunctions.

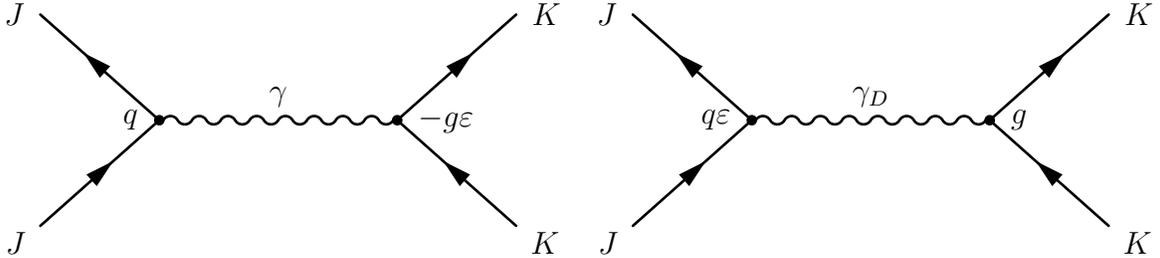
\begin{figure}[t]
\centering
\begin{fmffile}{EMschan1}
\begin{fmfgraph*}(180,80)
\fmfpen{1.0}
\fmfstraight
\fmfleft{i1,p1,i2}\fmfright{o1,p2,o2}
\fmfv{l= $J$}{i1}\fmfv{l=$K$}{o1}
\fmfv{l= $J$,l.a=180}{i2}\fmfv{l=$K$,l.a=0}{o2}
\fmf{fermion,tension=1}{i1,v1,i2}
\fmf{fermion,tension=1}{o1,v2,o2}
\fmf{boson, tension=1.0,label.side=left,label=$\gamma$,label.dist=5}{v1,v2}
\fmfv{decor.shape=circle,decor.filled=full,decor.size=1.5thick,l=$q$,l.a=180,l.d=8}{v1}
\fmfv{decor.shape=circle,decor.filled=full,decor.size=1.5thick,l=$-g\varepsilon$,l.a=0,l.d=8}{v2}
\end{fmfgraph*}
\end{fmffile}\hspace{1cm}
\begin{fmffile}{EMschan2}
\begin{fmfgraph*}(180,80)
\fmfpen{1.0}
\fmfstraight
\fmfleft{i1,p1,i2}\fmfright{o1,p2,o2}
\fmfv{l= $J$}{i1}\fmfv{l=$K$}{o1}
\fmfv{l= $J$,l.a=180}{i2}\fmfv{l=$K$,l.a=0}{o2}
\fmf{fermion,tension=1}{i1,v1,i2}
\fmf{fermion,tension=1}{o1,v2,o2}
\fmf{boson, tension=1.0,label.side=left,label=$\gamma_D$,label.dist=5}{v1,v2}
\fmfv{decor.shape=circle,decor.filled=full,decor.size=1.5thick,l=$q\varepsilon$,l.a=180,l.d=8}{v1}
\fmfv{decor.shape=circle,decor.filled=full,decor.size=1.5thick,l=$g$,l.a=0,l.d=8}{v2}
\end{fmfgraph*}
\end{fmffile}
\caption{\label{f.schan} $s$-channel production of magnetic monopoles. There are contributions from both the massless photon and massive dark photon.}
\end{figure}

As with scattering, there are diagrams from both the massless photon and the massive dark photon, see Fig.~\ref{f.schan}. Again, the opposite couplings in these two diagrams cause the simple massless photon pole to be replaced by
\beq
\frac{1}{p^4}\to\frac{m_D^4}{p^4\left(p^2-m_D^2 \right)^2}~,
\eeq
in the squared amplitude, where $p$ is the momentum flowing through the $s$-channel photon. This means that at energies, $E$, far above the dark photon mass this process is suppressed by $m_D^4/E^4$. Clearly, monopoles with mass $M$ have the largest production cross-section for $2M<E<m_D$. In this case the monopoles are tightly bound by the confining force, and the constituents are relativistic. We expect the binding energy to raise the mass of the bound state to $\sim m_D$ and for bound state effects to play a significant role in the cross section. If, on the other hand, $M\gg m_D$ the largest cross sections are near threshold, where we also produce a confined state, but a quirky one~\cite{Kang:2008ea,Terning:2019bhg}. In this limit we can use non-relativistic approximations in the rest frame of the bound state.

In either case, single photon production should dominate, as the kinetic mixing parameter $\varepsilon$ is small, so the magnetic coupling is perturbative. For this particular production channel discrete symmetries are particularly powerful. When two electric particles annihilate to produce a photon they only have local couplings to $A_\mu$, which has $J^{PC}=1^{--}$. From Table~\ref{dLJ} it is clear that only the fermionic ${}^3\! S_1$ and $^3\!D_1$ states and the scalar ${}^1\! P_1$ state have the appropriate quantum numbers to produce, or be produced by, a single photon. 

However, for the magnetic case the fermionic ${}^3\! S_1$ and $ ^3\!D_1$ states and the scalar $^1\!P_1$ state have $J^{PC}$ of $1^{+-}$, the same as $B_\mu$, which magnetic particles couple to locally. Consequently, single photon pair production of magnetic scalars is forbidden by $CP$ conservation as there is no $1^{--}$ magnetic scalar bound state. This confirms the conclusion of ref.~\cite{Ignatiev:1997pm} that single photon pair-production of scalar monopoles vanishes in $CP$ preserving theories. What has not been previously appreciated is that single photon production of magnetic fermions by electric particles also vanishes. This can be seen from the fact that while the magnetic fermion ${}^1\! P_1$ state is a $1^{--}$, and so its production is allowed by $CP$ invariance, the magnetic particles have no \emph{single photon} coupling to this state. Consequently, if $CP$ is a good symmetry the amplitude for a single photon created by the annihilation of electric particles to produce a magnetic particle--antiparticle pair vanishes.

While the amplitude for single photon production of this type is always zero, the fermionic and scalar cases are distinct. In the scalar case the amplitude is forbidden by $CP$ invariance. Because of this we cannot see the cancellation of the spurious pole in the this case. When fermions are involved we do actually find that the pole cancels. This occurs because the production amplitude is allowed by discrete symmetries, and it is only the nature of the particle couplings that forces the amplitude to vanish. We demonstrate this cancellation explicitly in what follows.

We begin with electric fermions  with charges $\pm qe$ and mass $m$ which produce scalar magnetic monopoles with charges $\pm g4\pi/e$ through a single photon. In this case the amplitude is
\beq
\mathcal{M}=\frac{8\pi g \,m_D^2 }{p^2(p^2-m_D^2)\left(n\cdot p \right)}\varepsilon_{\mu\nu\alpha\beta}n^\alpha p^\beta \overline{v}(k_-)\gamma_\mu u(k_+) q^\nu~.
\eeq
where $k^\mu$ is the relative momentum between the electric particles and $q^\mu$ is the relative momentum between the magnetic particles. The photon carries the sum of the two particle momenta $p^\mu$. That is, we take the two electric particles to have momenta $k_\pm\equiv p^\mu/2\pm k^\mu$ and similarly the magnetic  particles to have momenta $q_\pm\equiv p^\mu/2\pm q^\mu$. 

There is only one configuration of the electric fermions that $CP$ invariance allows to connect to the spin-one magnetic scalar state. In particular, the $^1\! P_1$ fermionic bound state is a $1^{+-}$ and has the right quantum numbers to produce the magnetic scalars. We focus on this state by rewriting
\beq
\overline{v}(k_-)\gamma_\mu u(k_+)=\text{Tr}\left\{ \gamma_\mu u(k_+)\overline{v}(k_-)\right\}~,
\eeq
and using the identity derived in Appendix~\ref{a.RelProj}:
\beq
\left. u(k_+)\overline{v}(k_-)\right|_\text{singlet}=\frac{1}{2\sqrt{2}}\gamma^5\left[p\mathbb{I}_4-\frac{2m}{p}\slashed{p} +\frac{2}{p}\slashed{k}\slashed{p}\right]~,
\eeq
to project onto the spin singlet state in the center of momentum frame. One immediately sees that this projection vanishes, which is expected seeing as this state has no overlap with the vector current. However, we can use a Gordon identity
\beq
\overline{v}(k_-)\gamma^\mu u(k_+)=\overline{v}(k_-)\left(\frac{k^\mu}{m}-\frac{i\sigma^{\mu\nu}}{2m}p_\nu \right)u(k_+),
\eeq
where $\sigma^{\mu\nu}=\frac{i}{2}\left(\gamma^\mu\gamma^\nu-\gamma^\nu\gamma^\mu \right)$, to rewrite the form of the fermionic current. By associating the Levi-Civita tensor with the fermion bilinear and using the identity $\sigma^{\mu\nu}=\frac{i}{2}\varepsilon^{\mu\nu\alpha\beta}\sigma_{\alpha\beta}\gamma^5$, we find
\begin{align}
&\overline{v}(k_-)\gamma^\mu u(k_+)\varepsilon_{\mu\nu\alpha\beta}n^\alpha p^\beta=\nonumber\\
&\frac{1}{2m}\overline{v}(k_-)\left(2\varepsilon_{\mu\nu\alpha\beta}k^\mu n^\alpha p^\beta-p_\nu\sigma_{\alpha\beta} n^\alpha p^\beta\gamma^5-\left(n\cdot p \right)p^\beta\sigma_{\beta \nu}\gamma^5-p^2n^\alpha\sigma_{\nu\alpha}\gamma^5 \right) u(k_+).
\end{align}
Because the magnetic current is conserved $p\cdot K=0$ and we can drop the second term. We also use the relation
\beq
\overline{v}(k_-)p^\beta\sigma_{\beta\nu}\gamma^5u(k_+)=\overline{v}(k_-)i\gamma^5\left(k_{+\nu}-k_{-\nu} \right)u(k_+),
\eeq
to find
\begin{align}
&\overline{v}(k_-)\gamma^\mu u(k_+)\varepsilon_{\mu\nu\alpha\beta}n^\alpha p^\beta=\frac{1}{m}\overline{v}(k_-)\left(\varepsilon_{\mu\nu\alpha\beta}k^\mu n^\alpha p^\beta-i\left(n\cdot p \right)k_\nu\gamma^5+\frac{p^2}{2}n^\alpha\sigma_{\alpha\nu}\gamma^5 \right) u(k_+).\label{e.ExpandedOpk}
\end{align}

The term in \eqref{e.ExpandedOpk} with the Levi-Civita tensor vanishes in the trace, having no spinor matrix structure. The remaining terms both have one factor of $\gamma^5$ which combines with the same term in the projection matrix to give the identity. Within the projection matrix the term with a single gamma matrix does not contribute to the trace, since it leads to terms with odd numbers of gamma matrices. The remaining terms are
\begin{align}
&\frac{1}{2m\sqrt{2}}\text{Tr}\left\{\left[ -i\left(n\cdot p \right)q_\mu+\frac{p^2}{2}n^\alpha\sigma_{\alpha\mu}\right]\left[p\mathbb{I}_4+\frac{2}{p}\slashed{k}\slashed{p}\right] \right\}\nonumber\\
=&\frac{\sqrt{2}}{m}\left\{  -ip\left(n\cdot p \right)k_\mu+i\frac{p^2}{2}\frac{2}{p}\left[k_\mu(n\cdot p)-p_\mu(n\cdot k) \right]\right\}~.
\end{align}
The first two terms cancel, while the last vanishes when contracted with the conserved current. In short, we have
\beq
\text{Tr}\left\{\gamma^\mu u(k_+)\overline{v}(k_-)\varepsilon_{\mu\nu\alpha\beta}n^\alpha p^\beta\right\}=\frac{i\sqrt{2}p\left(n\cdot p\right)}{m}\left(k_\nu-k_\nu \right)=0\times\left(n\cdot p\right)~.
\eeq
This is still zero, as it had to be, but also comes with a factor of $(n\cdot p)$, cancelling the spurious pole. Thus we see explicitly that the electric vector current has no overlap with a $1^{+-}$ state, but that this vanishing result is unambiguous. That is, picking a configuration in which $n\cdot p=0$ does not affect the result.

Finally, we turn to single photon production of a fermions with magnetic charge $\pm g$ and mass $M$. The  amplitude is
\beq
\mathcal{M}=\frac{4\pi g \,m_D^2 }{p^2(p^2-m_D^2)\left(n\cdot p \right)}\varepsilon_{\mu\nu\alpha\beta}n^\alpha p^\beta\overline{v}(k_-)\gamma_\mu u(k_+)\overline{u}(q_+)\gamma_\nu v(q_-).
\eeq
In this case there are two configurations allowed by $CP$. The first is nearly identical to the previous case. We project the incoming electric charges onto the $1^{+-}$ state that matches the magnetic vector current. This amplitude vanishes as above, but the spurious pole is cancelled. In the second case the electric particles are in the $1^{--}$ state and we must project the magnetic state onto the ${}^1P_1$ state with  $J^{PC}=1^{--}$. This amplitude also vanishes, but can be rewritten so that the cancellation of the spurious pole is explicit.

The Gordon decomposition
\beq
\overline{u}(q_+)\gamma^\mu v(q_-)=\overline{u}(q_+)\left(\frac{q^\mu}{M}+\frac{i\sigma^{\mu\nu}}{2M}p_\nu \right)v(q_+),
\eeq
allows us to write 
\begin{align}
&\overline{u}(q_+)\gamma^\nu v(q_-)\varepsilon_{\mu\nu\alpha\beta}n^\alpha p^\beta=\frac{1}{M}\overline{u}(q_+)\left(\varepsilon_{\mu\nu\alpha\beta}q^\nu n^\alpha p^\beta+i\left(n\cdot p \right)q_\mu\gamma^5+\frac{p^2}{2}n^\alpha\sigma_{\alpha\mu}\gamma^5 \right) v(q_-).\label{e.ExpandedOp}
\end{align}
We project onto the spin singlet configuration using the result from Appendix~\ref{a.RelProj}
\beq
\left.\text{Tr}\left\{\mathcal{O}v(q_-)\overline{u}(q_+) \right\}\right|_\text{singlet}=-\frac{1}{2\sqrt{2}}\text{Tr}\left\{\mathcal{O}\gamma^5\left[p\mathbb{I}_4+\frac{2M}{p}\slashed{p} -\frac{2}{p}\slashed{q}\slashed{p}\right]\right\}~,
\eeq
and find
\begin{align}
\text{Tr}\left\{\gamma^\nu v(q_-)\overline{u}(q_+)\varepsilon_{\mu\nu\alpha\beta}n^\alpha p^\beta\right\}=&\frac{i\sqrt{2}p\left(n\cdot p\right)}{M}\left(q_\nu-q_\nu \right)=0\times\left(n\cdot p\right)~.
\end{align}
Thus, the amplitudes allowed by conservation laws have no spurious pole, though they also happen to be zero.

It is easy to see how running these processes in reverse also works out. If we had magnetic scalars or fermions producing an electron pair, then  the monopoles in the $1^{+-}$ state would produce a  magnetic photon which has the right quantum numbers to produce the electric $^1P_1$ state. However, the electric current has a zero matrix element in this state, so the amplitude vanishes, although we can again explicitly see the cancelation of the spurious pole.

\subsection{Photon Fusion}
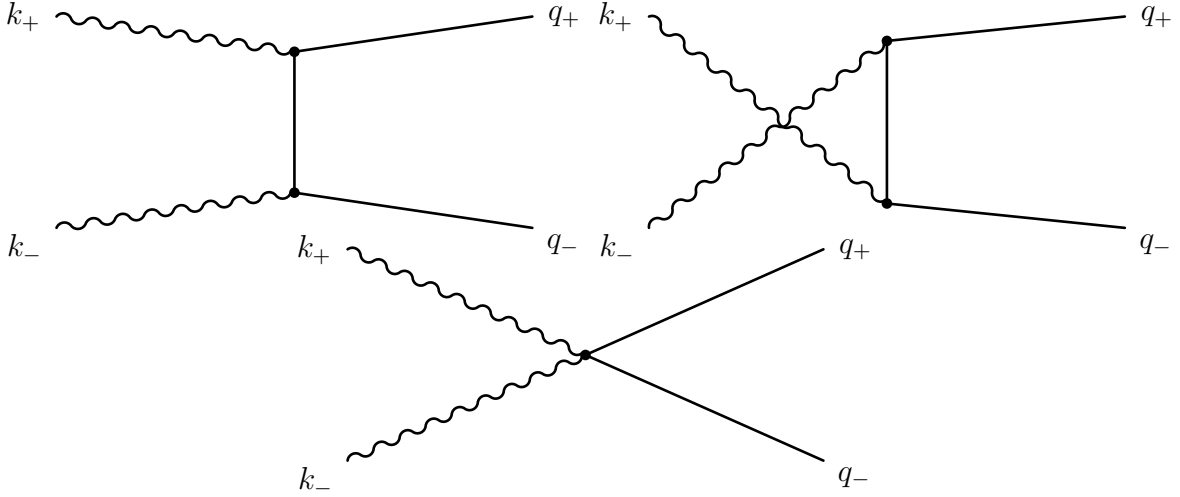
\begin{figure}[t]
\centering
\begin{fmffile}{PhotFusScal1}
\begin{fmfgraph*}(180,80)
\fmfpen{1.0}
\fmfstraight
\fmfleft{i1,p1,i2}\fmfright{o1,p2,o2}
\fmfv{l= $k_-$}{i1}\fmfv{l=$q_-$}{o1}
\fmfv{l= $k_+$,l.a=180}{i2}\fmfv{l=$q_+$,l.a=0}{o2}
\fmf{photon,tension=1}{i1,v1}\fmf{plain,tension=1}{v1,o1}
\fmf{photon,tension=1}{i2,v2}\fmf{plain,tension=1}{v2,o2}
\fmf{plain,tension=0.5}{v1,v2}
\fmfv{decor.shape=circle,decor.filled=full,decor.size=1.5thick}{v1}
\fmfv{decor.shape=circle,decor.filled=full,decor.size=1.5thick}{v2}
\end{fmfgraph*}
\end{fmffile}\hspace{1cm}
\begin{fmffile}{PhotFusScal2}
\begin{fmfgraph*}(180,80)
\fmfpen{1.0}
\fmfstraight
\fmfleft{i1,p1,i2}\fmfright{o1,p2,o2}
\fmfv{l= $k_-$}{i1}\fmfv{l=$q_-$}{o1}
\fmfv{l= $k_+$,l.a=180}{i2}\fmfv{l=$q_+$,l.a=0}{o2}
\fmf{phantom,tension=1}{i1,v1,o1}
\fmf{phantom,tension=1}{i2,v2,o2}
\fmf{photon,tension=0.1}{i1,v2}\fmf{plain,tension=0.1}{v2,o2}
\fmf{photon,tension=0.1}{i2,v1}\fmf{plain,tension=0.1}{v1,o1}
\fmf{plain,tension=0.2}{v1,v2}
\fmfv{decor.shape=circle,decor.filled=full,decor.size=1.5thick}{v1}
\fmfv{decor.shape=circle,decor.filled=full,decor.size=1.5thick}{v2}
\end{fmfgraph*}
\end{fmffile}\\[0.25cm]
\begin{fmffile}{PhotFusScal3}
\begin{fmfgraph*}(180,80)
\fmfpen{1.0}
\fmfstraight
\fmfleft{i1,p1,i2}\fmfright{o1,p2,o2}
\fmfv{l= $k_-$}{i1}\fmfv{l=$q_-$}{o1}
\fmfv{l= $k_+$,l.a=180}{i2}\fmfv{l=$q_+$,l.a=0}{o2}
\fmf{photon,tension=1}{i1,v1}\fmf{plain,tension=1}{v1,o1}
\fmf{photon,tension=1}{i2,v1}\fmf{plain,tension=1}{v1,o2}
\fmfv{decor.shape=circle,decor.filled=full,decor.size=1.5thick}{v1}
\end{fmfgraph*}
\end{fmffile}
\caption{\label{f.scalPhotFus} Photon fusion of scalar monopoles}
\end{figure}

Clearly, pair production of magnetic monopoles must proceed through two or more photons. The simplest process to check is two photons fusing to create two monopoles of mass $M$ and charge $\pm g$. For scalar monopoles there are three diagrams which contribute, see Fig.~\ref{f.scalPhotFus}. For simplicity we consider two on-shell photons with momenta $k_\pm^\mu=p^\mu/2\pm k^\mu$ creating two magnetic scalars with momenta $q_\pm^\mu=p^\mu/2\pm q^\mu$. The amplitude is simply
\beq
\mathcal{M}=2g^2\tilde{\epsilon}^\mu(k_+)\tilde{\epsilon}^\nu(k_-)\left[g_{\mu\nu}-\frac{q_{\nu-}q_{\mu+}}{k_-\cdot q_-} -\frac{q_{\mu-}q_{\nu+}}{k_+\cdot q_-} \right]~,
\eeq
where the $\tilde{\epsilon}_\mu$ are polarization vectors corresponding to $B_\mu$. 

Using the relation in Eq.~\eqref{e.polarizationRelation} that the $A_\mu$ and $B_\mu$ polarization vectors are related by
\beq
\tilde{\epsilon}^\nu(k)=-\frac{\varepsilon^{\nu\mu\alpha\beta}n_\mu k_\alpha \epsilon_\beta(k)}{n\cdot k}~,
\eeq
we can consider the amplitude which is related to photon fusion due to photon emission  from electric particles. Of course, this also introduces two spurious poles, and in general they do not cancel in the amplitude. However, when we considered  $t$-channel scattering we saw that the poles cancelled if we took the vector $n^\mu$ to lie along the vector that points from one magnetic particle to the other, i.e. along the physical string. This is equivalent to taking $n^\mu\propto q^\mu$, the relative momentum between the monopoles. In this case, using the notation $\epsilon^\mu(\pm)=\epsilon^\mu(k_\pm)$, we find
\begin{align}
\mathcal{M}=&\frac{32g^2}{p^4-16(k\cdot q)^2}\left\{\epsilon(+)\cdot\epsilon(-) \left[\frac{p^4}{16}+\frac{p^2q^2}{2}+(k\cdot q)^2 \right]+k\cdot\epsilon(+) \,k\cdot\epsilon(-) \left(\frac{p^2}{2}+4q^2 \right) \right.\nonumber\\
&\left.-\frac{p^2}{2}\, q\cdot \epsilon(+)\, q\cdot \epsilon(-) -2 k\cdot q \left[ k\cdot \epsilon(+)\,  q\cdot \epsilon(-)+q\cdot \epsilon(+)\, k\cdot \epsilon(-) \right]\right\}~.
\end{align}
 Here, the spurious poles in $n\cdot k$ have all canceled, thus confirming that $n^\mu$ corresponds to the physical  string joining the two magnetic particles.
 
 A similar, if considerably more tedious, calculation can be completed for fermionic monopoles. There are two diagrams, which are similar in form to the first two diagrams of Fig.~\ref{f.scalPhotFus}. The cancelation of the spurious poles is not obvious from the amplitude, but upon considering the squared amplitude and summing over fermion spins we find (using {\sc Feyncalc}~\cite{Shtabovenko:2020gxv})
 \beq
 \sum_\text{spin}|\mathcal{M}|^2=16g^4\frac{p^8+8M^2p^6-32M^4p^4-128M^2p^2\left( k\cdot q\right)^2-256\left( k\cdot q\right)^4}{\left[p^4-16\left( k\cdot q\right)^2\right]^2}~.
 \eeq
 We again see that if $n^\mu\propto q^\mu$ then the spurious $n\cdot k$ poles cancel. 
 

\section{Conclusion\label{s.conc}}
Manifestly local Lagrangian theories with ``mutually non-local" charges have a two-point function that contains a spurious pole. When Dirac charge quantization holds, these poles need only cancel when amplitudes are calculated to all-orders in perturbation theory. In the soft photon  limit  the resummation has been performed and it was indeed found that the poles exponentiate into unobservable a topological phase \cite{Terning:2018udc}.  However, with kinetic mixing between a massless photon and a massive dark photon, charge quantization is lost and perturbative electric-magnetic interactions can result. In such theories the spurious poles must cancel order-by-order.

We have shown that the spurious poles are cancelled when the total magnetic charge in a given process is zero, as required by magnetic charge confinement. In the case of electric particles scattering off a confined monopole--anti-monopole pair, only states or transitions with a non-zero dipole moment contribute. For these states the relative position vector between  the monopole and anti-monopole has a non-zero expectation value. This means that the interactions of the electric particle with the two magnetic particles differ by a sign and a relative phase proportional to the separation. Including this effect is sufficient to demonstrate that the spurious pole cancels in $t$-channel scattering.
By taking careful account of discrete symmetries we were able to show that for pair production of magnetic particles  the spurious poles also cancel in physically allowed processes. We also demonstrated that single photon production of magnetic monopole pairs vanishes. 

It is interesting to compare the two very different situations with and without charge quantization.  An essential ingredient in the all-orders/topological calculation was the requirement that the world line of the monopole could be taken as a closed loop in spacetime.  With the usual inclusion of the ``point at infinity" this is of course possible.  In the case of the order-by-order cancellation the essential ingredient was the presence of both a monopole and a nearby anti-monopole. Considering the worldlines of the monopole and anti-monopole, and including the ``point at infinity", we again have a closed loop in spacetime. In the soft photon limit we are still sensitive to the second magnetic charge no matter how far away it is. For low-energy/non-relativistic interactions of monopoles confined with anti-monopoles photon exchange is still sensitive to both magnetic charges.  However, for a sufficiently hard photon we generally expect to be able to find some form of factorization so that we only need to consider scattering of individual constituents of the bound state. However the standard approach to the $S$-matrix breaks down in this case, as has been shown recently in ref.~\cite{Csaki:2020inw}, since there is  angular momentum stored in the Coulomb fields of the particles even when they are asymptotically far apart. This situation can be handled using on-shell methods when Dirac charge quantization holds \cite{Csaki:2020inw}, since then the angular momentum is automatically quantized correctly as well.  In the perturbative case where charge quantization fails to hold however the local notion of angular momentum is not quantized and it may be impossible to prove factorization in these theories. It will be interesting to see how on-shell methods can be applied to this case.

\appendix
\section*{Acknowledgments}
We thank Hsin-Chia Cheng, Csaba~Cs\'aki, Can Kilic, Markus Luty, Maxim Pospelov, Arvind Rajaraman, Yuri Shirman, and Ofri Telem for clarifying discussions. J.T.  is supported in part by the DOE under grant DE-SC-0009999. C.B.V is supported in part by NSF Grant No. PHY-1915005 and in part by Simons Investigator Award \#376204.


\section{Propagators in Two Potential Formulations\label{a.twoPot}}
In this appendix we calculate the propagators in the Zwanziger and Schwarz-Sen two potential formalisms using the path integral formulation. Beginning with Zwanziger's  Lagrangian we have
\beq
Z[J,K]=\int\mathcal{D}A\,\mathcal{D}B\exp\left\{i\int \frac{d^4k}{(2\pi)^4}\mathcal{L}_Z \right\},
\eeq
where the Lagrangian from Eq.~\eqref{e.zLag} is Fourier transformed into momentum space as
\begin{align}
\mathcal{L}_Z=&-\frac{1}{2e^2n^2}A_{\mu}(-k)\left[g^{\mu\nu}\left(n\cdot k \right)^2-\left(n\cdot k \right)\left(k^\mu n^\nu+k^\nu n^\mu \right)+k^2n^\mu n^\nu \right]A_\nu(k)\nonumber\\
&-\frac{1}{2e^2n^2}B_{\mu}(-k)\left[g^{\mu\nu}\left(n\cdot k \right)^2-\left(n\cdot k \right)\left(k^\mu n^\nu+k^\nu n^\mu \right)+k^2n^\mu n^\nu \right]B_\nu(k)\nonumber\\
+&\frac{1}{e^2n^2}A_\mu(-k)\left[ \left(n\cdot k \right)\varepsilon^{\mu\nu\alpha\beta}n_\alpha k_\beta\right]B_\nu(k)-J^\mu(p) A_\mu(-k)-\frac{4\pi}{e^2}K^\mu(k) B_\mu(-k)~.
\end{align}
To this we add the gauge fixing terms~\cite{Gubarev:1998ss}
\beq
\mathcal{L}_\text{G.F.}=\frac{1}{2e^2n^2}\left[\xi_A^2\left(n\cdot A \right)^2+\xi_B^2\left(n\cdot B \right)^2 \right]~,
\eeq
where $\xi_{A,B}$ are gauge fixing parameters with mass dimension one. Varying the Lagrangian with respect to these parameters enforces the axial-type gauges $A\cdot n=0$ and $B\cdot n=0$. Note, however, that these parameters need not be equal as each of $A_\mu$ and $B_\mu$ have an independent gauge invariance. 

The propagators are obtained by integrating over the $A_\mu$ and $B_\mu$ fields in the path integral. The  electric propagator $\Delta_{AA}$  then connects $J^\mu$ to $J^\mu$. Similarly we can obtain the $B_\mu$ to $B_\nu$ propagator as well as the $A_\mu$ to $B_\nu$ propagator. In what follows we should properly Wick rotate to Euclidean space before evaluating the functional integrals and then rotate back. This introduces no subtlety for this calculation, so we simply leave such steps implied. In ordinary QED we have
\begin{align}
\int \mathcal{D}A\exp\left\{i\int \frac{d^4k}{(2\pi)^4}-\frac{1}{2e^2} A_{\mu}{\mathcal K}^{\mu\nu}A_\nu+J^\mu A_\mu \right\}\nonumber\\
=\left( \text{Det}\, {\mathcal K}\right)^{-1/2}\exp\left\{i\int \frac{d^4k}{(2\pi)^4}\frac{e^2}{2}J^\mu {\mathcal K}^{-1}_{\mu\nu}J^\nu \right\}~.\label{e.GaussInt}
\end{align}
Factors like $\text{Det}\,M$ cancel in the normalization of the path integral, and the propagator is
\beq
\Delta_{\mu\nu} = -i \,{\mathcal K}^{-1}_{\mu\nu}~.
\eeq

With the inclusion of the gauge fixing terms the full path integral can be written as
\begin{align}
Z[J,K]=\int\mathcal{D}A\mathcal{D}B\exp&\left\{i\int \frac{d^4k}{(2\pi)^4}-\frac{1}{2e^2} A_{\mu}{\mathcal K}_{AA}^{\mu\nu}A_\nu-\frac{1}{2e^2} B_{\mu}{\mathcal K}_{BB}^{\mu\nu}B_\nu+\frac{1}{e^2}A_{\mu}{\mathcal K}_{AB}^{\mu\nu}B_\nu\right.\nonumber\\
&\phantom{AAA}\left.J^\mu A_\mu+\frac{4\pi}{e^2}K^\mu B_\mu \right\},\label{e.PathInt}
\end{align}
where
\begin{align}
{\mathcal K}_{AA}^{\mu\nu}=&\frac{1}{n^2}\left[g^{\mu\nu}\left(n\cdot k \right)^2-\left(n\cdot k \right)\left(k^\mu n^\nu+k^\nu n^\mu \right)+\left(k^2-\xi_A^2\right)n^\mu n^\nu\right] , \label{e.MAA}\\
{\mathcal K}_{BB}^{\mu\nu}=&\frac{1}{n^2}\left[g^{\mu\nu}\left(n\cdot k \right)^2-\left(n\cdot k \right)\left(k^\mu n^\nu+k^\nu n^\mu \right)+\left(k^2-\xi_B^2\right)n^\mu n^\nu\right] ,\label{e.MBB}
\end{align}
are symmetric matrices and
\beq
{\mathcal K}_{AB}^{\mu\nu}=\frac{n\cdot k}{n^2}\varepsilon^{\mu\nu\alpha\beta}n_\alpha k_\beta~,
\eeq
is an antisymmetric matrix in the tensor indices. Then, by using Eq.~\eqref{e.GaussInt} twice in succession we find
\begin{align}
Z[J,K]\propto&\int\mathcal{D}A\exp\left\{i\int \frac{d^4k}{(2\pi)^4}-\frac{1}{2e^2} A_{\mu}{\mathcal K}_{AA}^{\mu\nu}A_\nu-J^\mu A_\mu \right.\nonumber\\
&\phantom{AAAAAAAA}\left.+\frac{1}{2e^2}\left[4\pi K^\mu+A_\alpha {\mathcal K}^{\alpha\mu}_{AB} \right]{\mathcal K}^{-1}_{BB\mu\nu}\left[4\pi K^\nu- {\mathcal K}^{\nu\beta}_{AB}A_\beta \right]\right\}\nonumber\\
\propto&\int\mathcal{D}A\exp\left\{i\int \frac{d^4k}{(2\pi)^4}-\frac{1}{2e^2}A_\mu\left[{\mathcal K}_{AA}^{\mu\nu}+{\mathcal K}^{\mu\alpha}_{AB}{\mathcal K}^{-1}_{BB\alpha\beta}{\mathcal K}^{\beta\nu}_{AB} \right]A_\nu\right.\nonumber\\
&\left.\phantom{AAAA}-\left[J^\mu+\frac{4\pi}{e^2}K^\alpha {\mathcal K}^{-1}_{BB\alpha\beta}{\mathcal K}^{\beta\mu}_{AB} \right]A_\mu+\frac12\left(\frac{4\pi}{e^2}\right)^2K^\mu {\mathcal K}^{-1}_{BB\mu\nu}K^\nu\right\}\nonumber\\
\propto&\exp\left\{-\int \frac{d^4k}{(2\pi)^4} \frac{e^2}{2} J^\mu \Delta^{AA}_{\mu\nu}J^\nu- 4\pi\,J^\mu \Delta^{AA}_{\mu\delta}{\mathcal K}^{\delta\gamma}_{AB}{\mathcal K}^{-1}_{BB\gamma\nu}K^\nu\right.\nonumber\\
&\left. \phantom{AAA}-\frac{i}{2}\left(\frac{4\pi}{e}\right)^2K^\mu\left[{\mathcal K}^{-1}_{BB\mu\nu}-i{\mathcal K}^{-1}_{BB\mu\alpha}{\mathcal K}^{\alpha\beta}_{AB}\Delta^{AA}_{\beta\gamma}{\mathcal K}^{\gamma\delta}_{AB}{\mathcal K}^{-1}_{BB\delta\nu} \right]K^\nu\right\},
\end{align}
where in the last line we have used the notation
\beq
\Delta^{AA}_{\mu\nu}=-i\left[{\mathcal K}_{AA}^{\mu\nu}+{\mathcal K}^{\mu\alpha}_{AB}{\mathcal K}^{-1}_{BB\alpha\beta}{\mathcal K}^{\beta\nu}_{AB} \right]^{-1}.
\eeq
Of course, we could have integrated out the $A$ field first and would have found the magnetic currents coupled by
\beq
\Delta^{BB}_{\mu\nu}=-i\left[{\mathcal K}_{BB}^{\mu\nu}+{\mathcal K}^{\mu\alpha}_{AB}{\mathcal K}^{-1}_{AA\alpha\beta}{\mathcal K}^{\beta\nu}_{AB} \right]^{-1}.
\eeq
We can also define
\beq
\Delta^{AB}_{\mu\nu}=\Delta^{AA}_{\mu\alpha}{\mathcal K}^{\alpha\beta}_{AB}{\mathcal K}^{-1}={\mathcal K}^{-1}_{AA\mu\alpha}{\mathcal K}^{\alpha\beta}_{AB}\Delta_{BB\beta\nu}~.
\eeq
The inverses of the symmetric matrices can be found by noting that they must have the form
\beq
c_0\,g^{\mu\nu}+c_1\left(k^\mu n^\nu+k^\nu n^\mu \right)+c_2\, k^\mu k^\nu+c_3\,n^\mu n^\nu~.
\eeq
By simply requiring that some matrix ${\mathcal K}^{\mu\nu}$ and its inverse ${\mathcal K}^{-1}_{\mu\nu}$ satisfy ${\mathcal K}^{\mu\alpha}{\mathcal K}^{-1}_{\alpha\nu}=g^\mu_{\hphantom{\mu} \nu}$ the unknown coefficients can be determined. In this manner we find the usual results for the Zwanziger propagators~\cite{Gubarev:1998ss}
\begin{align}
\Delta_{\mu\nu}^{AA}=&-\frac{i}{k^2}\left[g_{\mu\nu}-\frac{k_\mu n_\nu+k_\nu n_\mu}{n\cdot k}-\frac{n^2(k^2-\xi_A^2)}{\xi_A^2\left(n\cdot k \right)^2}k_\mu k_\nu \right],\nonumber\\
\Delta_{\mu\nu}^{BB}=&-\frac{i}{k^2}\left[g_{\mu\nu}-\frac{k_\mu n_\nu+k_\nu n_\mu}{n\cdot k}-\frac{n^2(k^2-\xi_B^2)}{\xi_B^2\left(n\cdot k \right)^2}k_\mu k_\nu \right],\\
\Delta_{\mu\nu}^{AB}=&\frac{i}{k^2}\frac{\varepsilon_{\mu\nu\alpha\beta}n^\alpha k^\beta}{n\cdot k}~. \nonumber
\end{align}
We emphasize that the mixed propagator is completely independent of the gauge fixing parameters. 

From these results we can infer the relationship between the electric $\epsilon^\mu(k)$ and magnetic $\tilde{\epsilon}^\mu(k)$ polarization vectors. The numerators of the $AA$ and $BB$ propagators are identified with the sum over polarizations
\beq
\sum \epsilon_\mu\epsilon_\nu^{\ast}~, \ \ \ \ \sum \tilde{\epsilon}_\mu\,\tilde{\epsilon}^{\ast}_\nu~,
\eeq
while the mixed propagator numerator comes from the sum
\beq
\sum \epsilon_\mu\tilde{\epsilon}_\nu^{\ast}~.
\eeq
These are all reproduced if we have the relation~\cite{Colwell:2015wna,Strominger:2015bla}
\beq
\tilde{\epsilon}^\nu(k)=-\frac{\varepsilon^{\nu\mu\alpha\beta}n_{\mu}k_\alpha \epsilon_\beta(k)}{n\cdot k}~.\label{e.polarizationRelation}
\eeq

Now, suppose we introduce an electric mass term,
\beq
\frac{1}{2e^2} m_D^2A_\mu A^\mu~,
\eeq
where the factor of $e^2$ appears because of our normalization of $A_\mu$. We no longer need to introduce the gauge fixing parameter $\xi_A$, but we do need to keep $\xi_B$. This is because the two vector fields each have a separate gauge invariance. Consequently, if one omits $\xi_B$ then matrices like ${\mathcal K}_{BB}^{\mu\nu}$ have no inverse. Therefore, while ${\mathcal K}_{BB}^{\mu\nu}$ is still given by Eq.~\eqref{e.MBB} we find
\beq
{\mathcal K}_{AA}^{\mu\nu}=\frac{1}{e^2n^2}\left[g^{\mu\nu}\left(\left(n\cdot k \right)^2-n^2m_D^2\right)-\left(n\cdot k \right)\left(k^\mu n^\nu+k^\nu n^\mu \right)+k^2n^\mu n^\nu\right]~.
\eeq
The same process as in the massless case can then be employed. We find
\begin{align}
\Delta_{\mu\nu}^{AA}=-&\frac{i}{k^2-m_D^2}\left[g_{\mu\nu}-\frac{1}{m_D^2}k_\mu k_\nu \right], \nonumber \\
\Delta_{\mu\nu}^{BB}=-&\frac{i}{k^2-m_D^2}\left[g_{\mu\nu}-\frac{m_D^2}{\left(n\cdot k\right)^2}\left(n^2g_{\mu\nu}-n_\mu n_\nu \right) \right. \\
& \quad\quad\quad\quad\quad\quad  \left. -\frac{k_\mu n_\nu+k_\nu n_\mu}{n\cdot k}  -\frac{n^2(k^2-\xi_B^2-m_D^2)}{\xi_B^2\left(n\cdot k \right)^2}k_\mu k_\nu \right], \nonumber\\
\Delta_{\mu\nu}^{AB}=&\frac{i}{k^2-m_D^2}\frac{\varepsilon_{\mu\nu\alpha\beta}n^\alpha k^\beta}{n\cdot k}~,\nonumber
\end{align}
which verifies the results of~\cite{Balachandran:1974nw}. The mixed propagator is again independent of the $B_\mu$ gauge fixing. Also worth pointing out is the non-decoupling behavior in the $BB$ propagator as $m_D\to\infty$. Intuitively, this is because as $m_D$ gets large the binding between the magnetic charges increases. Thus, it is plausible that the $n\cdot k$ pole is related to the confining potential~\cite{Gubarev:1998ss,Chernodub:1999tv,Zakharov:1999gn}.

While in Zwanziger's analysis the vector $n^\mu$ is taken spacelike and associated with the Dirac string, another two potential formulation was constructed by Schwarz and Sen~\cite{Schwarz:1993vs}. Their action is
\beq
S=-\frac12\int d^4x\left(B^{(\alpha)i}\varepsilon_{\alpha\beta}E^{(\beta)i}+B^{(\alpha)i}B^{(\alpha)i} \right),
\eeq
where $\alpha,\beta$ take values $1$ and $2$ corresponding to the two potentials, and $i$ is a spatial index. Their construction preserves rotational symmetry, and so might be associated with a Zwanziger like Lagrangian but with $n^\mu=(1,0,0,0)$. We may construct the 4-covariant version of their action in our notation by making associations
\begin{align}
E^{(1)}_\mu= \frac{n^\nu}{|n|} F^A_{\nu\mu}, \ \ \ E^{(2)}_\mu= \frac{n^\nu}{|n|} F^B_{\nu\mu}, \ \ \ B^{(1)\mu}=\frac{n_\nu}{|n|}\varepsilon^{\nu\mu\alpha\beta}\partial_\alpha A_\beta, \ \ \ B^{(2)\mu}=\frac{n_\nu}{|n|}\varepsilon^{\nu\mu\alpha\beta}\partial_\alpha B_\beta~.
\end{align}
Then the Lagrangian takes the form
\beq
\mathcal{L}_{SS}=\mathcal{L}_Z+\frac{1}{4e^2}\left(F_{\mu\nu}^AF^{A\mu\nu}+F_{\mu\nu}^BF^{B\mu\nu} \right).
\eeq
This still has the same form as the path integral in Eq.~\eqref{e.PathInt}, but we need to find a different gauge fixing term. 
If we ignore this for the moment we find
\begin{align}
{\mathcal K}_{AA}^{\mu\nu}=&\frac{1}{e^2n^2}\left\{g^{\mu\nu}\left[\left(n\cdot k \right)^2-n^2k^2\right]-\left(n\cdot k \right)\left(k^\mu n^\nu+k^\nu n^\mu \right)+n^2k^\mu k^\nu+k^2n^\mu n^\nu\right\} , \label{e.MAAss}\\
{\mathcal K}_{BB}^{\mu\nu}=&\frac{1}{e^2n^2}\left\{g^{\mu\nu}\left[\left(n\cdot k \right)^2-n^2k^2\right]-\left(n\cdot k \right)\left(k^\mu n^\nu+k^\nu n^\mu \right)+n^2k^\mu k^\nu+k^2n^\mu n^\nu\right\}  .\label{e.MBBss}
\end{align}
These are orthogonal to both $k^\mu$ and $n^\mu$ so cannot be inverted as they are. However, as we also saw in the Zwanziger cases, the mixed propagator only depends on the inversion of the $g^{\mu\nu}$ pieces, the other components being projected out by the Levi-Civita terms. In short, it is insensitive to whatever gauge fixing is employed. Thus, even without choosing gauge fixing terms we find
\beq
\Delta_{\mu\nu}^{AB}=\frac{i\left(n\cdot k \right)}{k^2\left[ \left(n\cdot k \right)^2-n^2k^2\right]}\varepsilon_{\mu\nu\alpha\beta}n^\alpha k^\beta~.
\eeq
This agrees exactly with the propagator derived by Weinberg~\cite{Weinberg:1965rz}. Thus, we see that the difference in the Weinberg and Zwanziger normalizations is tied to the choice of taking $n^\mu$ to be timelike or spacelike, respectively. 

\section{Electromagnetic Duality in Scalar Matter\label{a.ScalDual}}
This appendix demonstrates the $SL(2,\mathbb{Z})$ covariance of the Zwanziger formalism. Including a $CP$ violating parameter, $\theta,$ in the usual holomorphic coupling
\beq
\tau=\frac{\theta}{2\pi}+\frac{4\pi i}{e^2}~,
\eeq 
we can write the Lagrangian in Eq.~\eqref{e.zLag} as \cite{Csaki:2010rv,Terning:2018lsv} 
\begin{align}
\mathcal{L}_Z=&-\frac{\text{Im}(\tau)}{32\pi}\left[F_+^{\mu\nu}\left(F^A_{\mu\nu}-iF^B_{\mu\nu} \right)+F_-^{\mu\nu}\left(F^A_{\mu\nu}+iF^B_{\mu\nu} \right)\right] \nonumber\\
&-\text{Re}\left[ \left( A-iB\right)\cdot\left( J+\tau K\right)\right]~,\label{manifest}
\end{align}
where 
\beq
F_{\pm\mu\nu}\equiv F_{\mu\nu}\pm i\,{}^\ast\! F_{\mu\nu} ~.
\eeq
The equation of motion is then
\beq
\frac{\text{Im}(\tau)}{4\pi}\partial_\nu\left(F^{\mu\nu}+i{}^\ast\! F^{\mu\nu} \right)=J^\mu+\tau K^\mu~.
\eeq

Under an $SL(2,\mathbb{Z})$ duality transformation \cite{Cardy1,Vafa,Witten2,Lozano} the currents are mapped to
\beq
J^\mu\to bK'^\mu+dJ'^\mu, \ \ \ \ K^\mu\to aK'^\mu+cJ'^\mu.
\eeq
where $a,b,c,d$ are integers with $ad-bc=1$. 
The gauge fields transform \cite{Csaki:2010rv,Terning:2018lsv} as
\beq
A_\mu-iB_\mu &\to& \frac{1}{c\tau+d}\left( A'_\mu-i B'_\mu\right)~,\\
F_{+\mu\nu}&\to& \frac{1}{c\tau^\ast+d}F^\prime_{+\mu\nu}~. 
\eeq

As an example, consider the case of a Dirac fermion field with electric charge $q$ and magnetic charge $g$. The kinetic term for such a field is
\beq
\mathcal{L}_\psi=\overline{\psi}\left( i\slashed{D}-m_\psi\right)\psi,
\eeq
where
\beq
D_\mu=\partial_\mu+iqA_\mu+ig\frac{4\pi}{e^2}B_\mu~.
\eeq
We find the electric and magnetic currents to be
\beq
J^\mu=-\frac{\delta \mathcal{L}_\psi}{\delta A_\mu}=q\overline{\psi}\gamma^\mu\psi~, \ \ \ \ \frac{4\pi}{e^2}K^\mu=-\frac{\delta \mathcal{L}_\psi}{\delta B_\mu}=\frac{4\pi}{e^2}g\overline{\psi}\gamma^\mu\psi~.
\eeq
We see that to include the Witten effect~\cite{Witten} we simply replace $q$ by $q+g\,\theta/(2\pi)$. Thus, the fermion interactions are
\beq
\mathcal{L}_\text{int}=\overline{\psi}\gamma^\mu\psi\left[A_\mu\left(q+g\frac{\theta}{2\pi} \right) +B_\mu g\frac{4\pi}{e^2}\right]=\overline{\psi}\gamma^\mu\psi\left[A_\mu\left(q+g\,\text{Re}
(\tau)\right) +B_\mu g\,\text{Im}(\tau)\right].\label{e.ferInt}
\eeq
By making the substitutions
\beq
A_\mu=\frac{A'_\mu\left(c\,\text{Re}(\tau)+d \right) -B'_\mu c\,\text{Im}(\tau)}{|c\tau+d|^2}, \ \ \ B_\mu=\frac{B'_\mu\left(c\,\text{Re}(\tau)+d \right) +A'_\mu c\,\text{Im}(\tau)}{|c\tau+d|^2}, 
\eeq
and 
\beq
q=d q' + bg'~, \ \ \ \ g=a g' + c q'~,
\eeq
we find
\beq
\mathcal{L}_\text{int}=\overline{\psi}\gamma^\mu\psi\left[A'_\mu\left(q'+g'\frac{ac\left(\text{Re}(\tau)^2+\text{Im}(\tau)^2 \right)+(ad+bc)\text{Re}(\tau)+bd}{|c\tau+d|^2}
\right) +B'_\mu g'\frac{\text{Im}(\tau)}{|c\tau+d|^2}\right],
\eeq
where we have used the identity $ad-bc=1$.

While it is not immediately obvious, from the interaction terms one finds that after the duality transformation the holomorphic coupling $\tau$ is replaced by 
\beq
\tau'=\frac{a\tau+b}{c\tau+d}~,
\eeq
which includes the transformation law for the imaginary part of $\tau$: 
\beq
\text{Im}(\tau')=\frac{\text{Im}(\tau)}{\left|c\tau+d \right|^2}~.
\eeq
This agrees with the transformation of the kinetic terms,
and the new Maxwell Equations are:
\beq
\frac{\text{Im}(\tau')}{4\pi}\partial_\nu\left(F'^{\mu\nu}+i{}^\ast\! F'^{\mu\nu} \right)=J'^\mu+\tau' K'^\mu~.
\eeq

This duality also holds for scalars with electric charge $q$ and magnetic charge $g$. In this case the kinetic term is
\begin{align}
\mathcal{L}_\phi=&\left|D_\mu\phi \right|^2=\left|\partial_\mu\phi \right|^2-i\left[\phi^\ast D_\mu\phi-\phi\left( D_\mu\phi\right)^\ast \right]\left( A^\mu q+B^\mu g\frac{4\pi}{e^2}\right)\\
=&\left|\partial_\mu\phi \right|^2-i\left(\phi^\ast \partial_\mu-\phi \partial_\mu\phi^\ast \right)\left( A^\mu q+B^\mu g\frac{4\pi}{e^2}\right)+|\phi|^2\left( A_\mu q+B_\mu g\frac{4\pi}{e^2}\right)\left( A^\mu q+B^\mu g\frac{4\pi}{e^2}\right).\nonumber
\end{align}
This determines the currents
\begin{align}
J^\mu=&-\frac{\delta \mathcal{L}_\psi}{\delta A_\mu}=qi \left(\phi^\ast \partial^\mu\phi-\phi \partial^\mu\phi^\ast \right)-2q|\phi|^2\left( A^\mu q+B^\mu g\frac{4\pi}{e^2}\right),\\
\frac{4\pi}{e^2}K^\mu=&-\frac{\delta \mathcal{L}_\psi}{\delta B_\mu}=\frac{4\pi}{e^2}gi \left(\phi^\ast \partial^\mu\phi-\phi \partial^\mu\phi^\ast \right)-\frac{8\pi}{e^2}g|\phi|^2\left( A^\mu q+B^\mu g\frac{4\pi}{e^2}\right).
\end{align}
Note that because the currents depend on $A^\mu$ and $B^\mu$ we cannot write the interaction terms in the form given by Eq.~\eqref{e.zLag}. Nevertheless, when we include the Witten effect the interaction Lagrangian has the form
\begin{align}
\mathcal{L}_\text{int}=&-i\left(\phi^\ast \partial_\mu-\phi \partial_\mu\phi^\ast \right)\left[ A^\mu \left(q+g\frac{\theta}{2\pi}\right)+B^\mu g\frac{4\pi}{e^2}\right]\nonumber\\
&+|\phi|^2\left[ A_\mu  \left(q+g\frac{\theta}{2\pi}\right)+B_\mu g\frac{4\pi}{e^2}\right]\left[ A^\mu  \left(q+g\frac{\theta}{2\pi}\right)+B^\mu g\frac{4\pi}{e^2}\right].
\end{align}
Since, like the fermionic interaction term in Eq.~\eqref{e.ferInt} the scalar interaction is a function only of the combination
\beq
A^\mu \left(q+g\frac{\theta}{2\pi}\right)+B^\mu g\frac{4\pi}{e^2},
\eeq
its form is preserved by the $SL(2,\mathbb{Z})$ duality transformation. Thus, charged scalars and charged fermions transform in the same way under electromagnetic duality.


\section{Relativistic Projector\label{a.RelProj}}
In this appendix we construct a relativistic projection matrix for the analysis of the production of excited magnetic bound states. We begin by defining the spinors
\beq
u^s(p)=\left( \begin{array}{c}
\sqrt{p\cdot\sigma}\,\bm{\xi}\\
\sqrt{p\cdot\overline{\sigma}}\,\bm{\xi}
\end{array}\right), \ \ \ \ v^{s}(p)=\left( \begin{array}{c}
\sqrt{p\cdot\sigma}\,\bm{\xi}\\
-\sqrt{p\cdot\overline{\sigma}}\,\bm{\xi}
\end{array}\right),
\eeq
where $\bm{\xi}$ is a two component numerical spinor while $\sigma^\mu=(\mathbb{I}_2,\sigma^i)$ and $\overline{\sigma}=(\mathbb{I}_2,-\sigma^i)$ use the usual Pauli matrices. We are interested in rewriting amplitudes as
\beq
\overline{v}(k_-)\mathcal{O}u(k_+)=\text{Tr}\left\{u(k_+) \overline{v}(k_-)\mathcal{O}\right\}~, \ \ \ \  \overline{u}(q_+)\mathcal{O}v(q_-)=\text{Tr}\left\{v(q_-) \overline{u}(q_+)\mathcal{O}\right\}
\eeq
using the outer product of the spinors as a projection operator on the interaction in question. In this case we define $k^\mu_\pm=p^\mu/2\pm k^\mu$, and $q^\mu_\pm=p^\mu/2\pm q^\mu$ where $p^\mu$ is the momentum of the bound state and $k^\mu$ or $q^\mu$ is the relative momentum of the constituents. The constituents are assumed to have the same mass, which, from for instance $k^2_+=k^2_-$, implies that $p\cdot k=p\cdot q=0$ and
\beq
m_k^2=\frac{p^2}{4}+k^2~, \ \ \ \ m_q^2=\frac{p^2}{4}+q^2~.
\eeq

We can immediately write
\beq
u(k_+) \overline{v}(k_-)=\left(\begin{array}{cc}
-\sqrt{k_+\cdot\sigma}\,\bm{\xi}\bm{\xi}^{\prime\dag}\sqrt{k_-\cdot\overline{\sigma}} & \sqrt{k_+\cdot\sigma}\,\bm{\xi}\bm{\xi}^{\prime\dag}\sqrt{k_-\cdot\sigma}  \\[0.2cm]
-\sqrt{k_+\cdot\overline{\sigma}}\,\bm{\xi}\bm{\xi}^{\prime\dag}\sqrt{k_-\cdot\overline{\sigma}} & \sqrt{k_+\cdot\overline{\sigma}}\,\bm{\xi}\bm{\xi}^{\prime\dag}\sqrt{k_-\cdot\sigma} 
\end{array} \right)~.\label{e.ProjMat}
\eeq
The two-component spinor outer product $\bm{\xi}\bm{\xi}^{\prime\dag}$ is evaluated according to the spins of the two particles in the desired state. For instance, for the singlet state and triplet states, respectively,
\beq
\bm{\xi}\bm{\xi}^{\prime\dag}=\frac{1}{\sqrt{2}}\mathbb{I}_2\;\;\text{Singlet}~, \ \ \ \ \ \bm{\xi}\bm{\xi}^{\prime\dag}=\frac{1}{\sqrt{2}}\bm{\epsilon}\cdot\bm{\sigma}\;\;\text{Triplet}~,
\eeq
where $\bm{\epsilon}$ is the polarization vector of the spin-1  state. 

The products like $\sqrt{V\cdot \sigma}$ are understood to mean the square-root of the eigenvalues of the matrix
\beq
V\cdot \sigma=\left( \begin{array}{cc}
V_0+V_3 & V_1-iV_2\\
V_1+iV_2 & V_0-V_3
\end{array}\right)~.
\eeq
Using the notation $\sqrt{\bm{V}\cdot\bm{V}}\equiv v$, this matrix is diagonalized by 
\beq
S_V=\frac{1}{\sqrt{2v(v+V_3)}}\left( \begin{array}{cc}
v+V_3 & V_1-iV_2\\
V_1+iV_2 & -v-V_3
\end{array}\right)~.
\eeq
It is straightforward to check that
\begin{align}
S_V \,V\cdot\sigma \,S_V&=\left( \begin{array}{cc}
V_0+v& 0\\
0 & V_0-v
\end{array}\right)=V_0\mathbb{I}_2+v\sigma_3~, \\
S_V \,V\cdot \overline{\sigma}\,S_V&=\left( \begin{array}{cc}
V_0-v& 0\\
0 & V_0+v
\end{array}\right)=V_0\mathbb{I}_2-v\sigma_3~,
\end{align}
and that $S_V=S_V^{-1}$.

Using the notation $a_\pm=\sqrt{V_0\pm v}$ we can write
\begin{align}
\sqrt{V\cdot\sigma}=&S_V\left( \begin{array}{cc}
a_+& 0\\
0 &a_-
\end{array}\right)S_V=\frac{1}{a_++a_-}\left(V\mathbb{I}_2+V\cdot\sigma \right),\\
\sqrt{V\cdot\overline{\sigma}}=&S_V\left( \begin{array}{cc}
a_-& 0\\
0 & a_+
\end{array}\right)S_V=\frac{1}{a_++a_-}\left(V\mathbb{I}_2+V\cdot\overline{\sigma} \right)~.
\end{align}
Using the properties of the Pauli matrices it is simple to check that
\beq
\sqrt{V\cdot\sigma}\sqrt{V\cdot\overline{\sigma}}=\mathbb{I}_2\sqrt{V_0^2-v^2}=\mathbb{I}_2V~,
\eeq
as expected. 

Now, we can use these results to determine how the massive spinors behave under boosts. At rest we have
\beq
u^s(0)=\sqrt{m}\left( \begin{array}{c}
\bm{\xi}\\
\bm{\xi}
\end{array}\right), \ \ \ \ v^{s}(0)=\sqrt{m}\left( \begin{array}{c}
\bm{\xi}\\
-\bm{\xi}
\end{array}\right),
\eeq
while for general momentum $p$ we have
\begin{align}
u^s(p)=&\frac{1}{a_++a_-}\left( \begin{array}{c}
(m+p\cdot\sigma)\bm{\xi}\\
(m+p\cdot\overline{\sigma})\bm{\xi}
\end{array}\right)=\frac{\sqrt{m}}{a_++a_-}\left[ \mathbb{I}_4+\frac{1}{m}\slashed{p}\gamma^0\right]u^s(0)~, \\
v^s(p)=&\frac{1}{a_++a_-}\left( \begin{array}{c}
(m+p\cdot\sigma)\bm{\xi}\\
-(m+p\cdot\overline{\sigma})\bm{\xi}
\end{array}\right)=\frac{\sqrt{m}}{a_++a_-}\left[ \mathbb{I}_4+\frac{1}{m}\slashed{p}\gamma^0\right]v^s(0)~.
\end{align}
We can also write 
\beq
u(0)\overline{v}(0)=m\left( \begin{array}{cc}
-\bm{\xi}\bm{\xi}^{\prime\dag} & \bm{\xi}\bm{\xi}^{\prime\dag}\\
-\bm{\xi}\bm{\xi}^{\prime\dag} & \bm{\xi}\bm{\xi}^{\prime\dag}
\end{array}\right)=m\gamma^5\left( \mathbb{I}_4-\gamma^0\right)\left( \begin{array}{cc}
\bm{\xi}\bm{\xi}^{\prime\dag} &0\\
0 & \bm{\xi}\bm{\xi}^{\prime\dag}
\end{array}\right)~,
\eeq
and
\beq
v(0)\overline{u}(0)=m\left( \begin{array}{cc}
\bm{\xi}\bm{\xi}^{\prime\dag} & \bm{\xi}\bm{\xi}^{\prime\dag}\\
-\bm{\xi}\bm{\xi}^{\prime\dag} & -\bm{\xi}\bm{\xi}^{\prime\dag}
\end{array}\right)=-m\gamma^5\left( \mathbb{I}_4+\gamma^0\right)\left( \begin{array}{cc}
\bm{\xi}\bm{\xi}^{\prime\dag} &0\\
0 & \bm{\xi}\bm{\xi}^{\prime\dag}
\end{array}\right)~.
\eeq

We are now ready to evaluate the entries of the projection matrix in Eq.~\eqref{e.ProjMat}. The momenta, $k_+^\mu$ and $k_-^\mu$, lead to a shorthand $a^\pm_{\pm}$ where the superscript labels the momentum. It is straightforward to determine that
\beq
a^+_++a^+_- =\sqrt{p_0+2k_0+2m}~, \ \ \ \ a^-_++a^-_- =\sqrt{p_0-2k_0+2m}~.
\eeq
 We then find
 \beq
 u(k_+)\overline{v}(k_-)=\frac{\gamma^5\left[ \mathbb{I}_4m+\slashed{k}_+\gamma^0\right]}{\sqrt{(p_0+2m)^2-4k_0^2}}\left( \mathbb{I}_4-\gamma^0\right)\left( \begin{array}{cc}
\bm{\xi}\bm{\xi}^{\prime\dag} &0\\
0 & \bm{\xi}\bm{\xi}^{\prime\dag}
\end{array}\right)\left[ \mathbb{I}_4m+\gamma^0\slashed{k}_-\right]~.
 \eeq
 For the singlet configuration we have
  \beq
 u(k_+)\overline{v}(k_-)=\frac{\gamma^5\left[ \mathbb{I}_4m+\slashed{k}_+\gamma^0\right]}{\sqrt{2}\sqrt{(p_0+2m)^2-4k_0^2}}\left( \mathbb{I}_4-\gamma^0\right)\left[ \mathbb{I}_4m+\gamma^0\slashed{k}_-\right]
 \eeq
We are interested in specific states described by $J^{PC}$ quantum numbers. However, parity is most simply defined in the center of momentum frame, or the rest frame of the system, so we consider the projector in that frame, where
\beq
p^\mu=(p,\vec{0}), \ \ \ \ k^\mu=(0,\vec{k})~.
\eeq
We can express the singlet projector in this frame as
\beq
 u(k_+)\overline{v}(k_-)=\frac{1}{2\sqrt{2}}\gamma^5\left[p\mathbb{I}_4-\frac{2m}{p}\slashed{p} +\frac{2}{p}\slashed{k}\slashed{p}\right]~.\label{e.SingProje}
\eeq
Similarly, we find the related singlet projector to be
\beq
 v(q_-)\overline{u}(q_+)=-\frac{1}{2\sqrt{2}}\gamma^5\left[p\mathbb{I}_4+\frac{2m}{p}\slashed{p} -\frac{2}{p}\slashed{q}\slashed{p}\right]~.\label{e.SingProjm}
\eeq

In this same frame we can express the triplet spin matrix as
\beq
\left( \begin{array}{cc}
\bm{\xi}\bm{\xi}^{\prime\dag} &0\\
0 & \bm{\xi}\bm{\xi}^{\prime\dag}
\end{array}\right)=\frac{1}{\sqrt{2}}\gamma^5\gamma^0\slashed{\epsilon}~,
\eeq
where $\epsilon^\mu(p)$ is the polarization vector of the spin-1 state. Note that it satisfies $\epsilon(p)\cdot p=0$. We then find the triplet configuration projectors to be
\begin{align}
 u(k_+)\overline{v}(k_-)=&\frac{1}{4\sqrt{2}(2m+p)}\left[\mathbb{I}_4(2m+p)+2\slashed{k}+\frac{2m+p}{p}\slashed{p}+\frac{2}{p}\slashed{k}\slashed{p} \right]\slashed{\epsilon}\left( 2\slashed{k}-\frac{2m+p}{p}\slashed{p}\right)~,\label{e.TripProje} \\
  v(q_-)\overline{u}(q_+)=&\frac{1}{4\sqrt{2}(2m+p)}\left[\mathbb{I}_4(2m+p)+2\slashed{q}-\frac{2m+p}{p}\slashed{p}-\frac{2}{p}\slashed{q}\slashed{p} \right]\slashed{\epsilon}\left( 2\slashed{q}+\frac{2m+p}{p}\slashed{p}\right) ~.\label{e.TripProjm}
\end{align}

\subsection{Electric Scattering Example}
As an example of how these results can be used we consider the familiar process of one species of electric particle annihilating and producing another species, $f_1\overline{f}_1\to f_2\overline{f}_2$. In the center of momentum frame with $f_1$ and $\overline{f}_1$ having momenta $p^\mu/2+k^\mu$ and $p^\mu/2-k^\mu$, respectively, while $f_2$ and $\overline{f}_2$ have momenta $p^\mu/2+q^\mu$ and $p^\mu/2-q^\mu$. The amplitude is simply
\beq
\mathcal{M}=\frac{e^2q_1q_2}{p^2}\overline{v}(k_-)\gamma^\mu u(k_+)\overline{u}(q_+)\gamma_\mu v(q_-)~,
\eeq
and the squared amplitude, when spins are summed over, is simply
\beq
\sum_\text{spin}\left|\mathcal{M}\right|^2=\frac{4e^4q_1^2q_2^2}{p^4}\left[16\left(k\cdot q\right)^2+p^2\left(p^2+4m_1^2+4m_2^2 \right) \right]~.\label{e.ElonePhoton}
\eeq

This amplitude can also be computed by writing
\beq
\mathcal{M}=\frac{e^2q_1q_2}{p^2}\text{Tr}\left\{\gamma^\mu u(k_+)\overline{v}(k_-)\right\}\text{Tr}\left\{\gamma_\mu v(q_-)\overline{u}(q_+)\right\}~,
\eeq
and using the singlet or triplet spin projections of the fermion--antifermion pairs. One can determine immediately, for instance, that the spin singlet configurations vanish:
\beq
\text{Tr}\left\{\gamma^\mu u(k_+)\overline{v}(k_-)\right\}=0, \ \ \ \text{Tr}\left\{\gamma_\mu v(q_-)\overline{u}(q_+)\right\}=0~.
\eeq
Conversely, for the triple configuration
\begin{align}
\text{Tr}\left\{\gamma^\mu u(k_+)\overline{v}(k_-)\right\}=&\frac{\sqrt{2}}{2m_1+p}\left[4k^\mu\left(k\cdot\epsilon_i \right)+(2m_1+p)p\epsilon_i^\mu \right] ~,\\
\text{Tr}\left\{\gamma^\mu v(q_-)\overline{u}(q_+)\right\}=&\frac{\sqrt{2}}{2m_2+p}\left[4q^\mu\left(q\cdot\epsilon_f \right)+(2m_2+p)p\epsilon_f^\mu \right] ~,
\end{align}
where the $\epsilon_{i,f}^\mu$ are the polarization vectors of the initial and final $J=1$ fermion configurations. While this makes the amplitude appear different, the squared amplitude summed over the spin polarizations leads again to Eq.~\eqref{e.ElonePhoton}. Writing the amplitude in this way we see explicitly the $S$-wave (without a factor of the relative momenta) and $D$-wave (with two factors of the relative momentum) contribute to the scattering amplitude, as expected from the bound state analysis.

\end{document}